\documentclass[a4paper,fleqn]{cas-dc}

\usepackage[numbers]{natbib}

\usepackage[figuresright]{rotating}
\usepackage{float}
\usepackage{amsmath,amssymb,amsfonts}
\usepackage{algorithmic}
\usepackage{graphicx}
\usepackage{textcomp}

\usepackage{listings}
\usepackage{color}
\usepackage{xspace}
\usepackage{xcolor}
\usepackage{float}%provide float environment
\usepackage{booktabs}%provide \toprule, \midrule, \bottomrule
\usepackage{caption}
\DeclareCaptionLabelSeparator{custom}{.}
\usepackage{cleveref}
\crefname{figure}{\textcolor{cyan}{Fig.}}{\textcolor{cyan}{Fig.}}
\crefname{table}{\textcolor{cyan}{Table}}{\textcolor{cyan}{Table}}
\crefname{section}{\textcolor{cyan}{Section}}{\textcolor{cyan}{Section}}
\usepackage{stfloats}%tables occupy two columns
\usepackage{url} %url support for reference
\usepackage{framed} %Add borders to a paragraph
\usepackage[T1]{fontenc} %make \textbf to bold
\usepackage{multirow}
\usepackage{multicol}
\usepackage[ruled,linesnumbered,noend]{algorithm2e}
\usepackage{threeparttable}
\usepackage{subfigure}
\usepackage{mathrsfs}
\usepackage{inconsolata}
\usepackage{pifont}
\usepackage{balance}
\usepackage{verbatim}
\usepackage{color, colortbl}
\usepackage{tikz}
\usepackage{makecell}
\usepackage{hyperref}
\usepackage{indentfirst}
\usepackage{ragged2e}
\usepackage[most]{tcolorbox}

\newcommand{\ourname}{\textsc{CatchAll}\xspace}
\newcommand{\dataset}{RepoExEval\xspace}
\newcommand{\datasetexec}{RepoExEval-Exec\xspace}

\definecolor{codeblue}{rgb}{0.1, 0.1, 0.6}
\definecolor{codecomment}{gray}{0.45}
\definecolor{placeholder}{rgb}{0.8, 0.4, 0.1}

\definecolor{codebg}{RGB}{248,248,248}   % 背景色
\definecolor{codeframe}{RGB}{200,200,200} % 边框色
\definecolor{keywordcolor}{RGB}{0,0,160}  % 关键词
\definecolor{commentcolor}{RGB}{0,120,0}  % 注释
\definecolor{stringcolor}{RGB}{163,21,21} % 字符串

\lstdefinestyle{custom}{
    backgroundcolor=\color{codebg},
    frame=single,
    rulecolor=\color{codeframe},
    basicstyle=\ttfamily\footnotesize,
    keywordstyle=\color{keywordcolor}\bfseries,
    commentstyle=\color{commentcolor}\itshape,
    stringstyle=\color{stringcolor},
    numberstyle=\tiny\color{gray},
    numbers=left,
    numbersep=5pt,
    tabsize=4,
    breaklines=true,
    showstringspaces=false,
    captionpos=b
}

\SetCommentSty{mycommfont}
\begin{document}
\let\WriteBookmarks\relax
\def\floatpagepagefraction{1}
\def\textpagefraction{.001}

% Short title
\shorttitle{\ourname: Repository-Aware Exception Handling with Knowledge-Guided LLMs}

% Short author
\shortauthors{Qingxiao Tao et~al.}

% Main title of the paper
\title [mode = title]{\ourname: Repository-Aware Exception Handling with Knowledge-Guided LLMs} 

\author[a]{Qingxiao Tao}
\ead{tao_qingxiao@sjtu.edu.cn}

\author[a]{Xiaodong Gu}
\ead{xiaodong.gu@sjtu.edu.cn}

\author[a]{Hao Zhong}
\ead{zhonghao@sjtu.edu.cn}

\author[a]{Beijun Shen\corref{cor1}}
\ead{bjshen@sjtu.edu.cn}
\cortext[cor1]{Corresponding Author: Beijun Shen}

\address[a]{School of Computer Science, Shanghai Jiao Tong University, Shanghai, China}

% Here goes the abstract
\begin{abstract}
Exception handling is a vital forward error-recovery mechanism in many programming languages, enabling developers to manage runtime anomalies through structured constructs (e.g., \texttt{try-catch} blocks). 
Improper or missing exception handling often leads to severe consequences, including system crashes and resource leaks. 
While large language models (LLMs) have demonstrated strong capabilities in code generation, they struggle with exception handling at the repository level, due to complex dependencies and contextual constraints.
In this work, we propose \ourname, a novel LLM-based approach for repository-aware exception handling.
\ourname equips LLMs with three complementary layers of exception-handling knowledge:
(1) \emph{API-level exception knowledge}, obtained from an empirically constructed API–exception mapping that characterizes the exception-throwing behaviors of APIs in real-world codebases;
(2) \emph{repository-level execution context}, which captures exception propagation by modeling contextual call traces around the target code; and
(3) \emph{cross-repository handling knowledge}, distilled from reusable exception-handling patterns mined from historical code across projects.
The knowledge is encoded into structured prompts to guide the LLM in generating accurate and context-aware exception-handling code.
To evaluate \ourname, we construct two new benchmarks for repository-aware exception handling: a large-scale dataset \dataset and an executable subset \datasetexec. Experiments demonstrate that \ourname consistently outperforms state-of-the-art baselines, achieving a CodeBLEU score of 0.31 (\emph{vs.} 0.27\% for the best baseline), intent prediction accuracy of 60.1\% (\emph{vs.} 48.0\%), and Pass@1 of 29\% (\emph{vs.} 25\%). These results affirm \ourname's effectiveness in real-world repository-level exception handling.
\end{abstract}

%Research highlights
\begin{highlights}
\item The first repository-aware approach for automated exception handling using large language models.
\item Integrates three levels of exception-handling knowledge, covering API-level exceptions, repository-level execution context, and cross-repository handling knowledge, to guide exception handling generation.
\item Automatically mines large-scale real-world code repositories to construct high-quality knowledge for exception handling.
\item Demonstrates superior performance in repository-aware exception handling tasks compared to state-of-the-art approaches.
\end{highlights}

% Keywords
% Each keyword is seperated by \sep

\maketitle

\section{Introduction}
\label{sec:intro}
%Why
Exception handling is a critical mechanism in modern programming languages, forming a cornerstone for developing robust and reliable software systems~\cite{5070548}. 
When a program encounters an abnormal runtime condition, the corresponding exception handling mechanism triggers predefined recovery actions. This design isolates the error-handling logic from regular program flow, enhancing the robustness, readability, and maintainability of software. Despite its importance, exception handling is frequently misunderstood, underused, or incorrectly implemented by developers~\cite{shah2010understanding, asaduzzaman2016developers, melo2019unveiling}. Several studies~\cite{Kery2016, 7961532} have shown that exception-related code often exhibits critical deficiencies, including incomplete implementations, inadequate test coverage, or complete omission, potentially leading to severe consequences such as system crashes, data corruption, and security vulnerabilities. These challenges drive the urgent need for automated techniques to assist developers in producing correct and maintainable exception-handling code~\cite{padua2017revisiting}.

%Related work
Early research in this domain focuses on rule-based techniques to detect exception-prone locations and synthesize \texttt{catch} blocks. These methods rely on manually defined heuristics, including method calls~\cite{acharya09:mining}, control flow paths~\cite{jo2004uncaught}, and API contracts~\cite{fetzer2004automatic}. While interpretable, such approaches struggle to handle rare or project-specific exceptions due to their rigidity and limited generalizability.
Recent work has adopted learning-based techniques~\cite{Zhong22a, ZhangWZSPL20} to enable data-driven, feature-agnostic exception handling.
These techniques leverage statistical patterns in source code, but still face challenges in capturing nuanced semantic and contextual information, especially for complex exception scenarios.

The advent of large language models (LLMs) has opened a new solution direction in automated exception handling. Due to the generalized code comprehension and generation capability of LLMs, they do not rely on manually-crafted heuristics to synthesize code or large-scale annotated datasets to train domain-specific models. Instead, researchers adopt knowledge prompt chaining techniques~\cite{KPC} and multi-agent systems~\cite{seeker} to guide LLMs in generating, verifying, and refining exception-handling code. This line of work has strong generalization across diverse codebases.

While LLMs achieve impressive results on simple code tasks, recent studies~\cite{liang2025can} report that their effectiveness degrades substantially when tasks require repository-level context. Existing LLMs primarily rely on internal parametric knowledge and often lack a deep understanding of exception semantics and repository-aware program behavior.
In modern software systems, exception handling is complicated by indirect API invocations, asynchronous callbacks, cross-method exception propagation, and other complex control-flow structures, all of which demand long-range dependency analysis and holistic reasoning across the codebase. These challenges highlight the necessity of capturing repository-level code semantics and exception mechanisms. Consequently, automatically acquiring repository-aware exception knowledge is critical for generating accurate and actionable exception-handling suggestions.

\begin{figure}[t]
\centering
\includegraphics[width=\columnwidth]{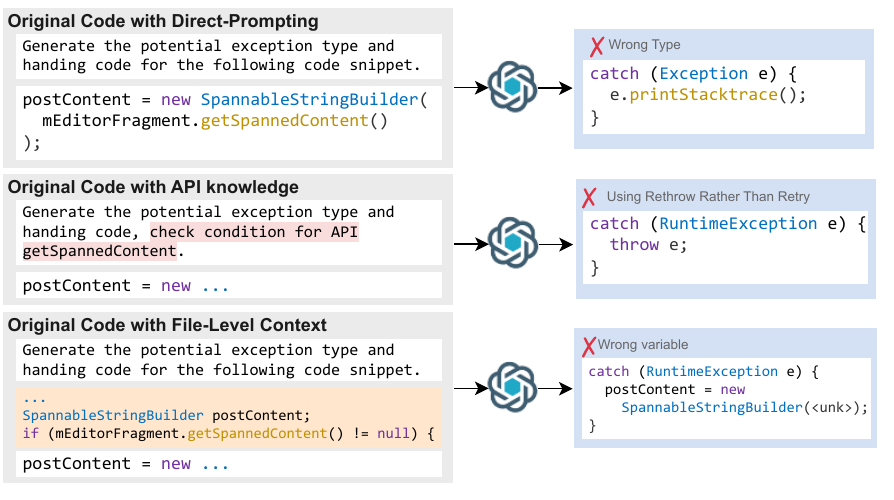}
\caption{A Motivating Example of Exception Handling in WordPress Android App. The exception originates from \texttt{TextView.getText()} deep in the call chain and is handled with fallback logic before causing application failure.}
\label{fig:motivation}
\end{figure}

We use a motivating example to illustrate the limitations of existing LLM-based generation approaches. Figure~\ref{fig:motivation} presents a representative exception-handling scenario from the Android project WordPress~\cite{wordpress_android_repo}. Although the failure is ultimately triggered by a \texttt{RuntimeException} thrown deep in the call chain at \texttt{TextView.getText()}, existing approaches fail in different ways. The direct prompting approach overlooks exception semantics and produces non-recoverable handling code. The API knowledge–driven prompt-chaining approach~\cite{KPC} enables the model to identify the correct exception type, but the generated handling logic degenerates into a naive rethrow without meaningful recovery. In contrast, the approach augmented with file-level context~\cite{ZhangWZSPL20} allows the model to infer a plausible handling intent, yet fails to correctly ground the handling logic in the concrete variables and execution context required by the surrounding code.

These limitations are mainly due to the lack of critical repository-level knowledge, including \emph{where} an exception originates along cross-method call traces, \emph{what} exception types are associated with specific APIs, and \emph{how} similar exceptions are effectively handled in real-world codebases. To address these challenges, we propose \ourname, a novel LLM-based approach for repository-aware exception handling.
\ourname equips LLMs with three layers of exception-handling knowledge:
(1) \emph{API-level exception knowledge}, which characterizes exception-prone behaviors of APIs through an empirically mined API-Exception mapping;
(2) \emph{repository-level execution context}, which captures exception propagation and control-flow dependencies via contextual call traces;
and (3) \emph{cross-repository handling knowledge}, which abstracts reusable exception-handling patterns from similar historical code examples.
\ourname is fully automated and extensible, requiring no manual design of templates or rules, and enables LLMs to generate exception-handling code that is explicitly aware of both API semantics and repository-specific execution context.

%Experimental results
%To evaluate the effectiveness of \ourname, we construct \dataset, a benchmark of 120,000 exception-handling code blocks  derived from popular open-source Android repositories on GitHub. We compare \ourname against five baselines, including Direct-Prompting, RepoCoder, ExAssist, Seeker, and KPC, based on two foundation models, GPT-4o and DeepSeek-V3. 
% Evaluation results demonstrate that \ourname outperforms all baseline methods by a significant margin in both CodeBLEU and intent prediction accuracy. 
%Evaluation results show that \ourname outperforms all baseline methods by 14.8$\sim$138.5\% in CodeBLEU and 25.2$\sim$230.2\% in intent prediction accuracy. 
%Specifically, under GPT-4o, \ourname achieves a CodeBLEU of 0.31 and an intent accuracy of 60.1\%, outperforming the best baseline by 4 points in CodeBLEU and over 12 points in intent accuracy.
% Ablation studies further emphasize the pivotal role of key components in our method.
%Removing contextual call traces, API-Exception mapping, or exception handling pattern leads to performance drops of 0.02–0.06 in CodeBLEU, confirming the necessity of each design choice.

The contributions of our work are summarized as follows:

\begin{itemize}

    \item We propose \ourname, the first repository-aware approach for automated exception handling, featuring a novel knowledge augmentation mechanism that integrates API-level exception knowledge, repository-level execution context, and cross-repository handling knowledge mined from large-scale code repositories.
    
    \item  We construct \dataset and \datasetexec to benchmark repository-aware exception handling. \dataset consists of 120K exception-handling code blocks collected from 3,200 open-source Android repositories on GitHub. \datasetexec comprises executable code from four representative repositories, in which generated \texttt{try-catch} blocks are integrated into the original codebase and validated via unit tests to assess functional correctness.
    
    \item  We conduct an extensive evaluation of \ourname against six representative methods, including Direct-Prompting, RepoCoder, ExAssist, Nexgen, Seeker, and KPC, under two LLM backbones, GPT-4o and DeepSeek-V3. The results show that \ourname consistently outperforms all baselines, yielding improvements of 14.8$\sim$138.5\% in CodeBLEU, 25.2$\sim$230.2\% in intent prediction accuracy, and 16$\sim$222.2\% in Pass@1.

\end{itemize}

\section{Related Work}
\label{sec:related}

\subsection{Empirical Studies on Exception Handling} 

Exception handling has been extensively studied through empirical research across multiple computing domains. 
The Java ecosystem has received particular attention. Cabral and Marques~\cite{cabral2007exception} conducted a comparative analysis of exception mechanisms between Java and .NET. Sena \emph{et al.}~\cite{sena2016understanding} examined exception patterns in Java libraries. Asaduzzaman \emph{et al.}~\cite{asaduzzaman2016developers} investigated exception usage in Java applications. ExChain~\cite{ExChain} analyzed exception-handling failures across large-scale Java systems.
Domain-specific analyses have significantly expanded this understanding. Koopman and DeVale~\cite{koopman2000exception} investigated exception handling in operating systems, Chen \emph{et al.}~\cite{chen2019understanding} analyzed exception-related bugs in cloud computing, and Bruntink \emph{et al.}~\cite{bruntink2006discovering} focused on embedded systems. Android platforms have been studied by both Coelho \emph{et al.}~\cite{coelho2017exception} and Fan \emph{et al.}~\cite{fan2018large} through large-scale empirical analyses.
Furthermore, Cacho \emph{et al.}~\cite{cacho2014does} and Anderson \emph{et al.}~\cite{10174045} made significant contributions by investigating exception-handling evolution across software versions.
%Shah \emph{et al.}~\cite{shah2010understanding} synthesized diverse exception handling viewpoints of Novices and Experts.

These empirical studies collectively reveal the critical role of exception handling in software reliability and maintainability. By examining exception usage patterns, failure causes, and real-world practices, they provide foundational insights that motivate the development of exception-handling techniques. %However, while these studies diagnose the challenges, they do not propose automated solutions for generating effective exception-handling code—highlighting the need for further research in this area.

%\subsection{Detecting and Repairing Exception-related Bugs}
\subsection{Automated Exception Handling}

As an important method to improve the reliability of software, automated exception handling \cite{SilvaVMGR24} has drawn a lot of attention in recent years. These approaches can be classified into rule-based approaches and learning-based approaches.

\textbf{Rule-based Approaches}. Rule-based approaches~\cite{ExChain, 9978172} leverage static analysis, control-flow inspection, and manually crafted heuristics to detect exception-prone locations and synthesize \texttt{catch} blocks. 
Acharya and Xie~\cite{acharya09:mining} mined method calls that should appear in \texttt{catch} clauses and Jana \emph{et al.}~\cite{jana2016automatically} extended their work with more advanced static analysis. Rahman~\cite{rahman2014use} recommended both method calls and detailed code snippets to handle exceptions.  Jo \emph{et al.}~\cite{jo2004uncaught} identified uncaught exceptions using control-flow analysis, while Fetzer \emph{et al.}~\cite{fetzer2004automatic} proposed generating wrappers for exception-prone methods. More recent rule-based efforts such as Nguyen \emph{et al.}~\cite{nguyen2019recommending,tam2020} predicted whether exceptions should be caught using hand-crafted code features, and ExAssist~\cite{ExAssist} applies fuzzy logic over mined historical patterns to recommend exception types and recovery APIs. ExceRef~\cite{ExceRef} advances this direction by providing automated refactoring of exception-handling structures using control-flow and dependency analysis, significantly reducing code risks and improving refactoring efficiency. Zhang \emph{et al.}~\cite{eha} designed an extended control-flow graph and symbolic execution framework to detect exception-handling bugs in C++ programs. %Although these approaches are interpretable and domain-specific, they often struggle with unseen code scenarios due to limited generalization capability.

\textbf{Learning-based Approaches}. Learning-based approaches ~\cite{9825827, 9462986} have become quite popular and powerful in recent years, since they are feature-agnostic and can automatically learn to handle exceptions from historical code samples without explicit definitions of detection rules or synthesis patterns.
Jia et al.~\cite{jia2021handle} trained a model to rank method calls based on their likelihood to throw exceptions, while Xu and Zhong~\cite{xu2021exception} identified inconsistencies between exception types and messages using classification models. ThEx~\cite{Zhong22a} further generalizes this by learning contextual features (\emph{e.g.}, throw location and variable names) to predict which exception should be thrown in a new context. Nexgen~\cite{ZhangWZSPL20} combines dual encoders and program slicing first to locate suitable \texttt{try} blocks and then synthesize corresponding \texttt{catch} clauses. CodeHunter~\cite{CodeHunter} applies a BERT-based model with Bi-LSTM to localize exception-prone code and predict exception types with strong empirical performance. Similarly, Neurex~\cite{Neurex} fine-tunes CodeBERT to detect \texttt{try-catch} boundaries, select catchable statements, and recommend exception types via multi-task learning, learning from repetitive exception-handling patterns across large codebases.%While these ML-based approaches improve adaptability, they usually require substantial task-specific training data and still fall short in capturing broader project-level semantics.
%Recently, LLMs have emerged as a powerful alternative for exception handling due to their ability to leverage large-scale pretraining and in-context learning. 

However, general-purpose LLMs without domain-specific exception knowledge struggle to effectively address complex real-world exception scenarios \cite{KPC, seeker}. 
To overcome this limitation, KPC~\cite{KPC} employs knowledge prompt chaining to iteratively verify and refine exception-handling code using detailed API documentation knowledge. 
Meanwhile, Seeker~\cite{seeker} develops a multi-agent collaboration framework to produce exception handling suggestions, supported by a curated exception handling knowledge base.
%RepoCoder~\cite{RepoCoder}, though not tailored to exception handling, demonstrates the effectiveness of incorporating repository-aware context into LLM-based code generation. 
%Compared to rule-based and ML-based techniques, LLMs offer several key advantages: they generalize well across diverse codebases without requiring manually curated features, and they reduce dependence on large annotated datasets by leveraging pretrained knowledge. However, most existing LLM-based approaches lack structured domain knowledge or deep integration with project-specific context, limiting their effectiveness in repository-scale exception handling.
Building upon these advances, \ourname achieves repository-aware exception handling by integrating three knowledge sources, \emph{i.e.}, exception-prone APIs, contextual call traces, and exception-handling patterns, all extracted by mining large-scale real-world code repositories.
%enabling the synthesis of accurate, coherent, and repository context-aware exception-handling code.

\section{Approach}
\label{sec:approach}

In this paper, we propose \ourname, a novel repository-level method for automated exception handling. \ourname enriches LLMs' awareness of repository context by integrating three key sources of knowledge: exception-prone APIs, execution context, and handling patterns derived from exemplars.

\subsection{Problem Formulation}

Given a code snippet within the current repository, exception handling aims to predict its exception type and wrap it with a \texttt{try-catch} block, yielding an exception-fortified code. This enhanced code must: (1) accurately capture potential runtime exceptions, (2) preserve semantic consistency with the repository context, and (3) comply with exception-handling conventions of the current project.

\subsection{Approach Overview}

\begin{figure*}[t]
\centerline{\includegraphics[width=0.95\textwidth, trim=0 -8 0 0, clip]{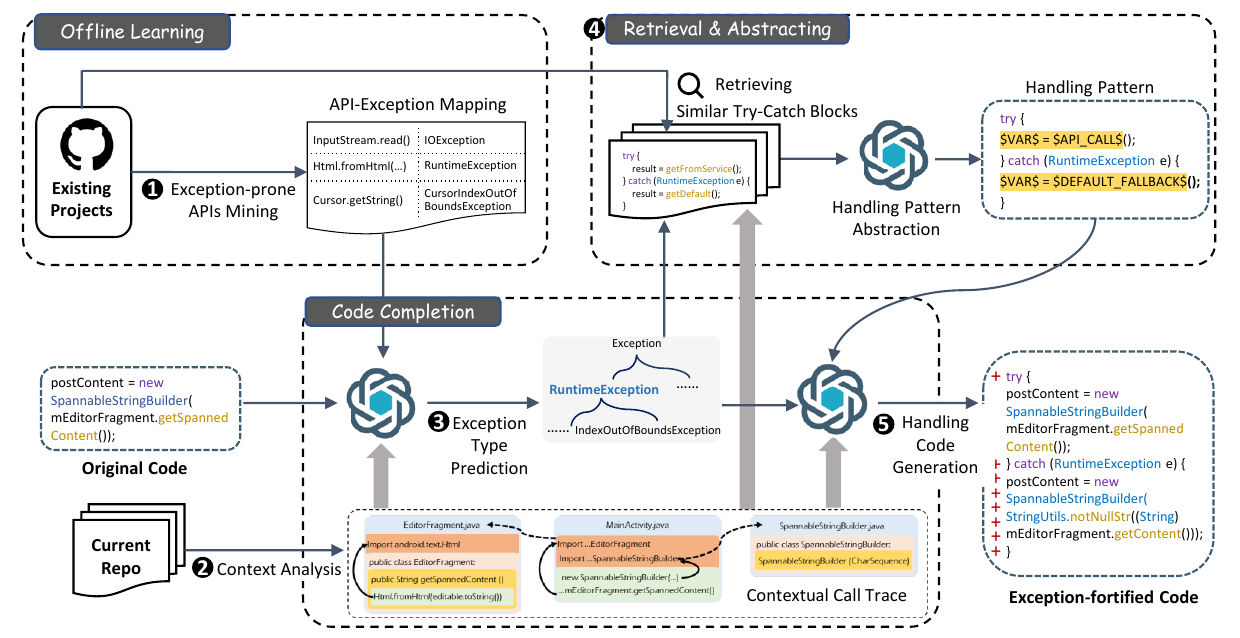}}
    \caption{Overview of \ourname.}
    \label{fig:overview}
\end{figure*}

Figure \ref{fig:overview} presents the overview of \ourname. The pipeline consists of five main steps, emulating developers' reasoning process. 
(1) First, to empower LLM with API knowledge, \ourname extracts exception-prone APIs from \texttt{try-catch} blocks in existing projects, establishing an API-exception mapping (Section~\ref{sec:Mapping}).
(2) For a given snippet, \ourname analyzes its execution context, namely, a repository-aware call trace to simulate exception propagation across function boundaries. This trace includes the upward call hierarchy from the try block's enclosing method, definitions of involved functions, their internal API calls, and import-based dependencies (Section~\ref{sec:ExtractContext}).
(3) Leveraging the exception-prone APIs and contextual call trace, \ourname predicts potential exception types that the snippet might throw. %\ourname queries the API-Exception mapping to identify potential exception type candidates. These candidates, along with the execution context, are processed by an LLM to predict the most likely exception type(s) that may be thrown in the input snippet 
(Section \ref{sec:PredictingType}).
(4) Based on the predicted exception types and function call stacks, \ourname retrieves relevant historical \texttt{try-catch} blocks from existing projects and abstracts them into a structured handling pattern (Section \ref{sec:RetrievalTemplating}). 
(5) Finally, \ourname instantiates the handling pattern, synthesizing context-sensitive exception-handling code tailored to the original code snippet (Section~\ref{sec:GeneratingHandler}). 

%By dynamically incorporating up-to-date exception knowledge, \ourname enhances the accuracy and adaptability of automated exception handling. 
% Each of the steps is elaborated in the following sections.

\subsection{Mining Exception-prone APIs}
\label{sec:Mapping}

To enhance LLMs' knowledge of common API usage,  \ourname automatically extracts exception-prone APIs from real-world codebases. It empirically models the relationships between API invocations and the exceptions they may throw within practical software systems. Our approach begins by automatically crawling numerous code repositories from GitHub and statically extracting their \texttt{try-catch} blocks using Tree-sitter\footnote{https://tree-sitter.github.io/tree-sitter/}, a robust and language-agnostic parser that provides accurate AST construction. For each \texttt{try} block, we apply the repository-aware context analysis algorithm (Algorithm~\ref{alg:context}) to derive the complete API usage trace. This trace includes not only the APIs directly within the \texttt{try} block but also those present in its entire call stack and surrounding methods.

Subsequently, we analyze all API invocations within the \texttt{try} block and its extended context, correlating them with the exception types captured in the corresponding \texttt{catch} block. We adopt a data-driven approach to define exception-prone APIs based on their empirical usage patterns in actual code. Specifically, any method invocation appearing within the context of a \texttt{try-catch} block, including its complete upstream call trace, is designated as exception-prone. The specific exception types caught in that context are then mapped to the API, forming a repository-grounded API-exception association. This mapping comprehensively covers standard libraries (e.g., JDK/Android SDK), third-party APIs, and user-defined methods. The coverage for a given repository is dynamically determined by the depth of our call-trace analysis, which is controlled by the parameter $d$ in Algorithm~\ref{alg:context}. This ensures the resulting mapping is specifically tailored to the codebase under analysis.

This process yields a many-to-many API-Exception mapping, where each API is linked to multiple exception types and vice versa. The resulting mapping is stored as structured domain knowledge, facilitating both exception type prediction and automated generation of exception-handling code.

\subsection{Analyzing Execution Context}

\begin{figure}[t]
\centerline{\includegraphics[width=\columnwidth, trim=0 0 0 0, clip]{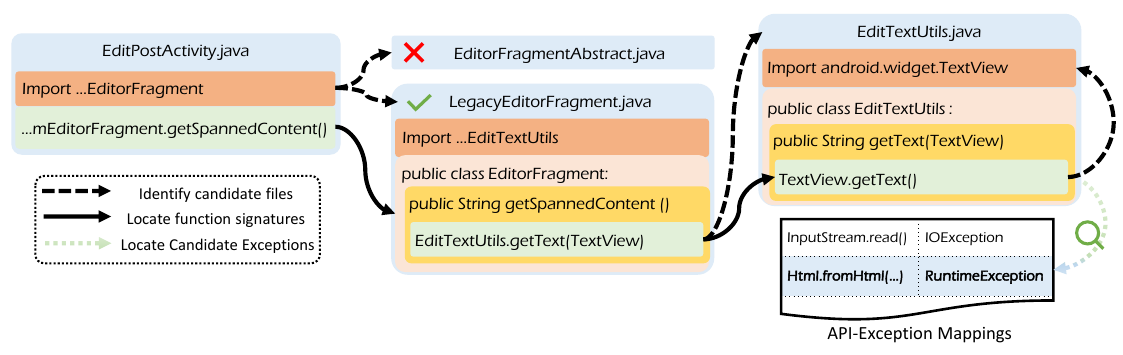}}
    \caption{An Example of a Contextual Call Trace Spanning Multiple Files.} 
    \label{fig:contextual}
\end{figure}

To improve the contextual understanding of exception causes and handling strategies, we extract a repository-aware execution context of the original code snippet. We recursively construct a multi-level call trace to simulate how exceptions propagate across function boundaries within the repository context. 
The trace spans multiple files and functions, reflecting how imported dependencies and cross-file invocations contribute to the runtime behavior relevant for exception handling, as illustrated in Figure~\ref{fig:contextual}.
The analysis specifically examines:

\label{sec:ExtractContext}

\begin{itemize}
    \item \textit{Imports}: All import statements declared in the current file.
    
    \item \textit{Call Hierarchy}: The complete call stack obtained by recursively tracing upward from the try-block's enclosing function through its caller hierarchy. Function resolution employs abstract syntax tree (AST) analysis combined with fuzzy name matching based on edit distance across both local definitions and imported symbols.        
  
    \item \textit{Function Signatures}: The signature information (e.g., function name, parameters) of all functions in the call stack.

    \item \textit{Filtered Function Bodies}: The bodies of functions within the call hierarchy, from which only invocation statements (i.e., function calls) are retained; the non-invocation code is discarded.
    
    \item \textit{Lexical Scope}: The enclosing class name (when applicable) and the immediate function name containing the \texttt{try} block.
\end{itemize}    

\begin{algorithm}[t]
\small
\caption{Contextual Call Trace Construction}
\label{alg:context}
\KwIn{Code segment $T$ in file $F$ within repository $R_T$}
\KwOut{Execution context $C_T$}

$C_T \leftarrow \emptyset$; \\
$C_T=C_T\cup F.import$ // Extract all import statements from $F$ and store in $C_T$; \\
Find the enclosing class and function of $T$ in $F$ and record their names in $C_T$; \\
Extract all API call statements in $T$ and initialize a stack $S$; \\

\While{$S$ is not empty and depth $\leq d$}{
    $f$ = Pop($S$); // Pop a function call from the stack \\
    Identify candidate files in $R_T$ using import names from $C_T$; \\
    Locate the definition(s) of $f$ from those files; \\
    \ForEach{definition $d$ of $f$}{
        Extract all function calls within $d$; \\
        Append $d$ and its calls to $C_T$; \\
        Push the new calls to $S$;
    }
}

\Return $C_T$;
\end{algorithm}

Algorithm~\ref{alg:context} outlines the procedure for constructing the execution context.
Given a code snippet $T$ in file $F$ within the current repository $R_T$, it first initializes an empty context object (line 1) and extracts static structural attributes from $F$, including import dependencies, the enclosing class name, and the method name (lines 2–3). It then identifies all API call statements within $T$ and initializes a call stack $S$ to enable repository-level analysis (line 4).
The core of the algorithm is a depth-bounded traversal (up to a maximum depth $d$) over the function call graph. In each iteration, a function call $f$ is popped from the stack $S$ (line 6). The algorithm locates the function signature(s) of $f$ through identifying candidate files in repository $R_T$ based on the import names collected in $C_T$, and analyzing these files to find actual function definitions (line 7 - 8). For each resolved definition $d$, \ourname extracts internal function calls (line 10), appends the definition and its callees to the context $C_T$ (line 11), and pushes the new calls onto the stack $S$ (line 12).
The traversal continues until the stack is empty or the maximum call depth is reached.

The extracted contextual call traces simulate repository-level execution behavior by resolving function calls across files using import relationships. This enables the capture of representative API usage patterns across the whole project, thereby enhancing the language model's generalization capability for exception handling in real-world repositories.

\subsection{Predicting Exception Types}
\label{sec:PredictingType}

With the collected exception-prone APIs and the execution context, we predict exception types for the input code snippet. 
We adopt a standard retrieval-augmented generation (RAG) framework for exception type prediction. The RAG framework includes two phases --- a retrieval phase and a generation phase.

In the retrieval phase, \ourname first extracts all API calls that appear throughout the call trace of the snippet. These API signatures are then used as search queries against our pre-built API–Exception mapping database, retrieving the complete set of potential exception types that the identified APIs may throw during execution.

In the generation phase, \ourname employs the LLM to predict the top-$k$ most likely exceptions using a zero-shot prompt that encodes: (1) the code snippet, (2) its contextual call trace, and (3) the candidate exception set. 
This prediction prompt encourages the model to accurately rank exceptions by their relevance to the given code context. By constraining the prediction to the candidate set, our method balances prediction accuracy and coverage, subsequently guiding the model to generate more precise, context-aware exception-handling code.

\begin{tcolorbox}[
    breakable,
    colframe=black, 
    colback=white, 
    coltitle=white, 
    colbacktitle=black, 
    title=Prompt for Predicting Exception Types,
    boxrule=0.8pt,
    fonttitle=\mdseries\small, 
    fontupper=\ttfamily\footnotesize, 
    sharp corners
]
\label{prompt1}
\textbf{\textcolor{codeblue}{System Prompt:}}\\
You are an expert in Java exception handling. Given a code snippet, a list of possible exception types, and the call trace, your task is to select the most appropriate exception type from the candidates.  
Output only a single, well-formed \texttt{catch} block that includes the selected exception type and a placeholder for the handling logic.

\vspace{0.2em}
\textbf{\textcolor{codeblue}{User Prompt:}}

{// Exception Type Candidates:} \\
\textcolor{violet}{\$exceptions\_list\$} 

{// Code snippet with surrounding context:} \\
\textcolor{violet}{\$code\_snippet\$} 

{/* Based on the provided code context and function calls, fill in the <mask0> token with a single appropriate exception type selected from the above candidates. The output should strictly follow this format: */} 

catch (<mask0> <mask1>) \{ \\
\hspace*{1.5em}<mask2> \\
\} 

%{\\ // Ensure that only one catch statement is returned, and do not add any extra explanation.}
\end{tcolorbox}

\subsection{Retrieval and Abstracting Prior Handling Samples}
\label{sec:RetrievalTemplating}
To generate robust and idiomatic exception-handling code, we augment the LLM with exemplary knowledge derived from prior handling samples.

\smallskip\subsubsection{Retrieving Similar Try-catch Blocks}
Given a code snippet in the current repository, its contextual call trace, and the predicted exception types, \ourname retrieves relevant historical \texttt{try-catch} blocks from existing codebases. We define a similarity metric of exception-handling code, considering three aspects:
\begin{itemize}
    \item \textit{Exception Type Similarity}: Computed using the hierarchical distance between the predicted exception type and that of the historical sample in the exception inheritance tree.
    \item \textit{API Sequence Similarity}: Calculated via Jaccard similarity between the sets of APIs used in both the original and historical code snippets.
    \item \textit{Try Block Similarity}: Measured by CodeBLEU~\cite{codebleu2020ren} to capture structural and semantic closeness.
\end{itemize}

These three scores are aggregated via weighted averaging. \ourname then retrieves the top-$K$ most similar \texttt{try-catch} blocks from existing projects. We empirically set $K = 20$ based on hyperparameter optimization in RQ4 (Section \ref{sec:RQ4}).

\smallskip\subsubsection{Abstracting Handling Patterns}
Prior works have discovered that limited code demonstrations may constrain model generalization~\cite{3608132}.
To address this limitation, we distill reusable exception-handling patterns from analogous historical code. The derived pattern aligns with the predicted behavior and execution context of the code snippet. 
We employ an LLM-driven approach to extract reusable exception-handling patterns from the top-$K$ retrieved \texttt{try-catch} blocks via a structured multi-phase abstraction process:

\begin{enumerate}
    \item \textit{Structural feature extraction}: Decompose each \texttt{catch} block into constituent elements including exception types, logging mechanisms, recovery procedures, cleanup operations, return value handling, and side effects.
    
    \item \textit{Pattern identification}: Analyze the samples to discover recurrent practices and isolate variable components.
    
    \item \textit{Pattern synthesis}: Generate a generalized pattern using parameterized placeholders (\emph{e.g.}, \verb|${{LogMethod}}|, \verb|${{RecoveryAction}}|).
    
    \item \textit{Pattern formalization}: Produce the final pattern in syntactically valid code format, ensuring readiness for subsequent exception-handling code generation.
\end{enumerate}

%The pattern abstraction prompt for exception handling is structured as follows:
\begin{tcolorbox}[breakable,
    colframe=black, 
    colback=white, 
    coltitle=white, 
    colbacktitle=black, 
    title=Prompt for Abstracting Exception-Handling Patterns,
    boxrule=0.8pt,
    fonttitle=\mdseries\small, 
    fontupper=\ttfamily\footnotesize, 
    sharp corners
]
\label{prompt2}
\textbf{\textcolor{codeblue}{System Prompt:}}\\
You are an experienced Java exception handling expert. Given a set of \texttt{try-catch} blocks, your goal is to analyze and abstract a general-purpose exception-handling pattern that captures the shared structure and variations across examples.

\vspace{0.2em}
\textbf{\textcolor{codeblue}{User Prompt:}}

Please follow these steps of reasoning:

\textbf{Step 1: Extraction features} \\
For each \texttt{catch} block, extract and summarize in a table:
\begin{itemize}
    \item Try-block AST
    \item Exception type
    \item Logging method
    \item Recovery or fallback logic
    \item ...
\end{itemize}

\vspace{0.3em}
\textbf{Step 2: Identify Patterns} \\
Determine shared components and varying parts among the blocks.

\vspace{0.3em}
\textbf{Step 3: Synthesize Patterns} \\
Design a unified template using placeholders such as \texttt{\$VAR\$}, \texttt{\$LOG\_METHOD\$}, \texttt{\$DEFAULT\_FALLBACK\$}, etc.

\vspace{0.3em}
\textbf{Step 4: Formalize Patterns} \\
Example format:
\begin{verbatim}
try {
    $VAR$ = $API_CALL$();
} catch (RuntimeException e) {
    $LOG_METHOD$(e);
    $VAR$ = $DEFAULT_FALLBACK$();
}
\end{verbatim}

\vspace{0.2em}
Input code samples:

\textcolor{violet}{\$try-catch\_blocks\$}
\end{tcolorbox}

%This abstraction process ensures that the resulting patterns capture exemplary knowledge while maintaining adaptability through well-defined slots.

\subsection{Generating Exception-Handling Code} 
\label{sec:GeneratingHandler}

\ourname instantiates the abstract handling pattern to generate context-aware exception-handling code tailored to the original code snippet.
The generation prompt incorporates the contextual call trace (including relevant API call sequences and surrounding code snippets), the predicted exception types, and the generalized handling pattern abstracted from similar historical cases. During instantiation, \ourname directs the language model to produce a \texttt{catch} block by filling placeholders in the pattern with context-aware content, thereby ensuring compliance with both exception-handling conventions and project-specific semantic constraints.

\begin{tcolorbox}[breakable,
    colframe=black, 
    colback=white, 
    coltitle=white, 
    colbacktitle=black, 
    title=Prompt for Generating Exception-Handling Code,
    boxrule=0.8pt,
    fonttitle=\mdseries\small, 
    fontupper=\ttfamily\footnotesize, 
    sharp corners
]
\label{prompt3}
\textbf{\textcolor{codeblue}{System Prompt:}}\\
You are an expert in Java exception handling that helps generate robust and context-aware Java \texttt{catch} blocks.  
Given a \texttt{try} block with exception type, its contextual call trace, and handling pattern,  
your job is to fill in the pattern and output only the appropriate \texttt{catch} block that handles the exception meaningfully.

\vspace{0.2em}
\textbf{\textcolor{codeblue}{User Prompt:}}

{// Referential exception handling pattern to follow when filling in the \texttt{catch} block:} \\
\textcolor{violet}{\$catch\_block\_pattern\$} 

{// Related API call sequence} \\
\textcolor{violet}{\$api\_call\_sequence\$} 

{// Relevant code fragments from the codebase} \\
\textcolor{violet}{\$relevant\_code\_fragments\$} 

{/* Based on the API call sequence, code context, and by following the above pattern style, fill in the above exception handling pattern to complete the following code: */} \\
try \{ \\
\hspace*{1.5em}\textcolor{violet}{\$try\_block\_code\$} \\
\} \\
catch (\textcolor{violet}{\$exception\_type\$} e) \{ 
\end{tcolorbox}

%Unlike prior methods that rely on rule-based patterns or example-driven substitution, our approach harnesses LLMs to synthesize semantically coherent and contextually adapted handling logic with superior flexibility and expressiveness.

\section{Experimental Setup}
\label{sec:setup}

We conduct experiments to evaluate the effectiveness of \ourname, aiming to answer the following research questions.
\begin{itemize}
\item \textbf{RQ1}: How does \ourname improve the state of the art? 

\item \textbf{RQ2}: How accurately are predicted exception types? 

\item \textbf{RQ3}: What are the impacts of individual components?

\item \textbf{RQ4}: What is the impact of the number of retrieved historical samples?

\end{itemize}

\subsection{Comparison Methods}
\label{sec:baselines}
We compare \ourname against four categories of exception-handling generation methods: rule-based completion (ExAssist), knowledge-enhanced reasoning (KPC), multi-agent collaboration (Seeker), and neural machine translation (Nexgen). 
Additionally, we adapt a state-of-the-art general repository-level code generation approach (RepoCoder) to exception handling. A direct LLM-based generation method without repository context is also included as the baseline.

Specifically, we evaluate \ourname against the following baseline methods:

\begin{itemize}
    \item \textbf{ExAssist}~\cite{ExAssist}: A rule-based method that learns fuzzy logic rules from thousands of high-quality programs to predict potential runtime exceptions and subsequently generates \texttt{try-catch} blocks to handle and recover from exceptions.

    \item \textbf{Nexgen}~\cite{ZhangWZSPL20}: A neural machine translation (NMT) based approach that utilizes an encoder-decoder architecture to localize exception-prone code and generate complete \texttt{catch} blocks. It is trained on a large-scale dataset of Java methods sourced from GitHub. For a fair comparison in our evaluation, we supply the model with the ground-truth location of the \texttt{try} block and task it solely with generating the \texttt{catch} block(s).
    
    \item \textbf{KPC}~\cite{KPC}: A knowledge-driven prompt chaining approach that leverages fine-grained exception-handling knowledge from API documentation to iteratively verify and rewrite generated code through targeted exception-handling prompts until all exceptions are properly addressed. 

    \item \textbf{Seeker}~\cite{seeker}: A multi-agent collaborative framework that orchestrates five specialized agents (Planner, Detector, Predator, Ranker, and Handler) to simulate expert reasoning, collaboratively infer exception-handling intents, and generate contextually appropriate handling code.

    \item \textbf{RepoCoder}~\cite{RepoCoder}: A general repository-level code completion method. To adapt it to our task, we place the input code inside a \texttt{try} block and leave the \texttt{catch} block empty, prompting RepoCoder to infer the exception type and generate corresponding handling logic. To ensure a fair comparison, we also implement RepoCoder-RG1, a single-round variant of RepoCoder that disables its iterative refinement mechanism, matching \ourname's one-round RAG strategy.

    \item \textbf{Direct-Prompting}: An LLM-based approach that generates exception-handling code from a standalone code snippet, independent of any repository-specific data or context.     
\end{itemize}

All baseline methods are implemented using their officially released code. 
For LLM-based methods such as \textit{RepoCoder}, \textit{KPC}, and \textit{Seeker}, we ensure fair comparison by embedding the same input code snippet into each method's original prompt construction pipeline, following their respective task setups and instructions. Consequently, each method receives the same functional context while preserving its own prompt design. Additionally, we replace their original backbone models with either GPT-4o\footnote{https://platform.openai.com/docs} or DeepSeek-V3\footnote{https://api.deepseek.com} to normalize model capabilities. %This allows us to attribute performance differences to the design of the methods themselves rather than the underlying model.

\subsection{Evaluation Metrics}
\label{sec:metrics}
We adopt three widely used metrics to  measure the performance of exception handling:

\textit{Pass@1}~\cite{chen2021evaluatinglargelanguagemodels}, which assesses the functional correctness of synthesized code by executing associated unit tests in a native project environment. Each generated try-catch block is integrated into its original project and run natively. A sample is considered correct if the test case designed to trigger the target exception passes as a result of the generated handler. Formally:
\begin{equation}
\text{Pass@1} = \frac{\#\text{ of correctly executed handlers}}{\#\text{ of total testable cases}}
\end{equation}

\textit{CodeBLEU}~\cite{codebleu2020ren}, a metric tailored for code generation that combines weighted n-gram matching, syntax-tree similarity, and semantic data-flow alignment, providing a more structure- and correctness-sensitive evaluation than vanilla BLEU.

\textit{IntentAcc} (Intent Prediction Accuracy)~\cite{IntentClassification}, which measures how accurately the predicted handling intents (\emph{e.g.}, logging, retry, error recovery, return, rethrow) match the ground truth. IntentAcc is computed as:

\begin{equation}
\text{IntentAcc} = \frac{1}{N} \sum_{i=1}^{N} \frac{|Y_i \cap \hat{Y}_i|}{|Y_i \cup \hat{Y}_i|}
\end{equation}
where $N$ is the number of exception-handling cases, and $Y_i$ and $\hat{Y}_i$ denote the sets of ground-truth and predicted intents, respectively.

\subsection{Dataset Construction}
\label{sec:datasets}

Due to the absence of publicly available datasets for repository-level exception handling, we create two novel benchmarks derived from real-world Android projects: \dataset and \datasetexec. 
\dataset is a large-scale benchmark collected from 3,200 repositories, designed to evaluate the accuracy of generated exception-handling code using CodeBLEU and IntentAcc metrics. 
The absence of unit tests and heavy dependencies prevents large-scale execution-based evaluation. To address this, we create \datasetexec, a compact executable benchmark from four repositories for assessing functional correctness using .

The dataset construction process comprises the following steps:

(1) \textit{Repository collection and filtering}. 
We first collect Android repositories from GitHub by searching for repositories containing both “Android” and “Java” as primary keywords, thereby capturing a broad spectrum of application domains and project structures. To ensure the selection of high-quality, authentic, and temporally valid projects, we apply a series of sequential filters:

\begin{itemize}
\item \textit{Code Quality Indicator}: We prioritize repositories with higher star counts as a proxy for code quality, community endorsement, and active maintenance.
\item \textit{Authenticity Verification}: Each repository is cross-referenced with the F-Droid open-source catalog to confirm its authenticity and to exclude abandoned, low-quality, or non-genuine Android projects.
\item \textit{Temporal Validity}: To prevent data contamination from Large Language Model pretraining corpora, we exclude projects updated after strict cutoff dates (September 2023 for GPT-4o and June 2024 for DeepSeek-V3).
\end{itemize}

This filtering yields a curated dataset of 3,200 high-quality, actively maintained Android repositories that balance domain representativeness with minimal inclusion of obsolete or low-quality projects. From these repositories, we extract 120,000 exception-handling code blocks, each containing the \texttt{try} and \texttt{catch} blocks, the raised exception types, associated call traces, and contextual metadata (such as enclosing class, method, file, and repository). To facilitate repository-aware reasoning, we further construct inter-procedural and repository-level contexts using Algorithm~\ref{alg:context}.

(2) \textit{\dataset construction}.
From the full collection, we randomly sample 1,000 exception-handling instances to form a benchmark test set for evaluation. The remaining samples are retained as a reference corpus to support API–exception mapping extraction and historical case retrieval within our method. All exception-handling intents (\emph{e.g.}, logging, retry, recovery, return, or throw) in the benchmark are manually annotated to ensure label accuracy. For the training and validation sets, intent labels are generated using rule-based heuristics and subsequently verified manually to guarantee quality.

(3) \textit{\datasetexec construction}.
To evaluate the functional correctness of generated exception-handling code, we construct \datasetexec, a small-scale executable benchmark comprising four projects: \textit{Aria2App}, \textit{openScale}, \textit{Overchan-Android}, and \textit{Signal-Android}. We select 100 real-world exception-handling instances from these projects, each accompanied by unit tests capable of triggering the corresponding exceptions. The generated catch blocks would be integrated into the original codebase and executed against these tests to assess runtime correctness and behavioral appropriateness.

Table~\ref{tab:dataset-stats} summarizes the statistics of the two benchmarks. In \dataset, the average project comprises 249 Java classes, and each exception-handling instance has an average call trace depth of 7.2 and spans 5.7 different files. \datasetexec exhibits greater complexity, with 665 classes per project, an average call trace depth of 7.7, and an average cross-file span of 6.1.

\begin{table}[t]
\caption{Statistics of the \dataset and the \datasetexec Benchmarks}
\label{tab:dataset-stats}
\centering
%\small
\resizebox{\columnwidth}{!}{
\begin{tabular}{lcc}
\toprule
\textbf{Metric} & \textbf{\dataset} & \textbf{\datasetexec} \\
\midrule
\# Repositories & 3,200 & 4 \\
\# Try-Catch Instances & 120,000 & 100 \\
\# Unit Tests & N/A & 174 \\
Avg. Call Trace Depth & 7.2 & 7.7 \\
Avg. Cross-File Span & 5.7 & 6.1 \\
Avg. Repository Size (\# Java Classes) & 249 & 665 \\
\bottomrule
\end{tabular}
}
\end{table}

\section{Results and Analysis}
\label{sec:results}

\subsection{Overall Performance (RQ1)}
\label{sec:RQ1}

\textbf{Setting.} 
We compare \ourname against various baseline methods for automated exception handling using the large-scale \dataset benchmark alongside the compact, executable \datasetexec. Our comparison includes both proprietary (GPT-4o) and open-source (DeepSeek-V3) state-of-the-art foundation models.

\begin{table}[t]
\centering
% \begin{threeparttable}
\centering
\caption{Performance of Various Repository-aware Exception Handling Approaches on \dataset}
\label{tab:overall_performance}
%\small
\resizebox{\columnwidth}{!}{
\begin{tabular}{llccccc}
\toprule
\textbf{Model} &\textbf{Approach} & \textbf{Pass@1} & \textbf{CodeBLEU} & \textbf{IntentAcc} \\
\midrule
- & ExAssist & 9\% & 0.13 & 33.9\% \\
 & Nexgen   & 14\% & 0.25 & 37.6\% \\
\midrule
GPT-4o & Direct-Prompting & 13\% & 0.24 & 46.8\% \\
  & RepoCoder RG1 & 21\% & 0.27 & 36.9\% \\
  & RepoCoder & 25\% & 0.27 & 48.0\% \\
  & KPC & 9\% & 0.13 & 18.2\% \\
  & Seeker & 15\% & 0.23 & 43.3\% \\
& \ourname (ours) & \textbf{29}\% & \textbf{0.31} & \textbf{60.1}\% \\
\midrule
DeepSeek-V3 & Direct-Prompting & 10\% & 0.24 & 43.4\% \\
  & RepoCoder RG1 & 21\% & 0.26 & 38.1\% \\
  & RepoCoder & 23\% & 0.27 & 47.3\% \\
  & KPC & 9\% & 0.13 & 18.2\% \\
  & Seeker & 15\% & 0.24 & 43.8\% \\
& \ourname (ours) & 26\% & \textbf{0.29} & \textbf{53.9}\% \\
\bottomrule
\end{tabular}
}
% \end{threeparttable}   
\end{table}

\begin{figure}[t]
    \centering
    \includegraphics[width=\columnwidth]{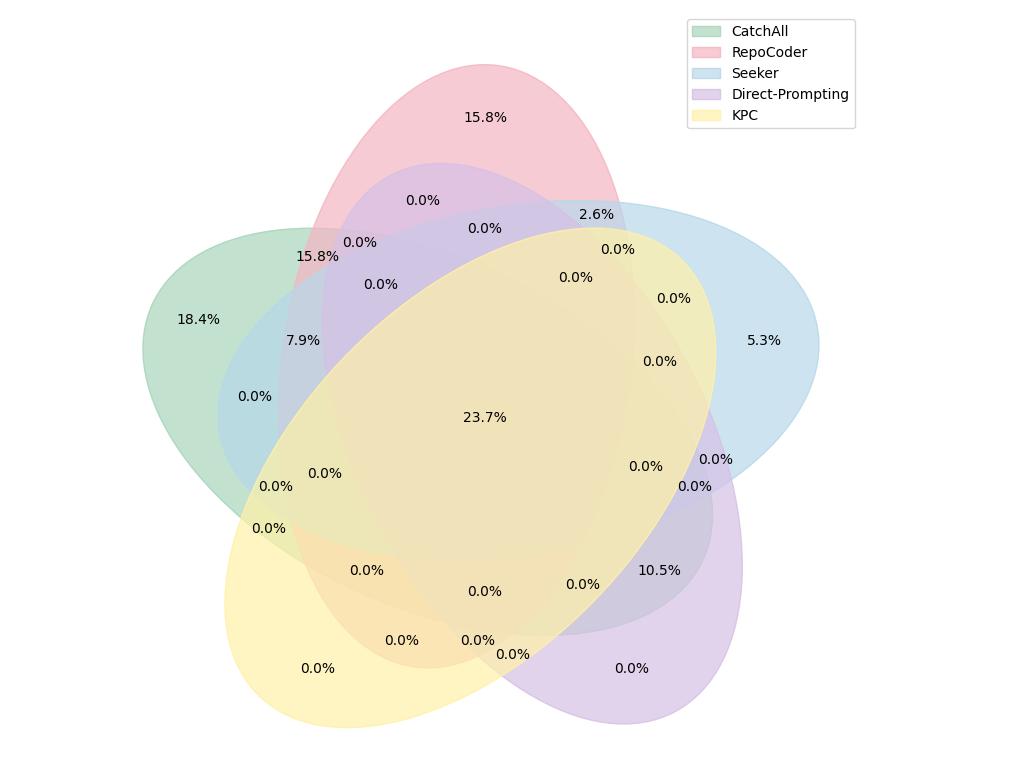}
    \caption{Venn Diagram of Correct Predictions across Different Methods under GPT-4o. 
    Each region represents the proportion of test instances correctly handled by a specific combination of methods.}
    \label{fig:rq1_venn}
\end{figure}

\smallskip\noindent\textbf{Result.} 
As shown in Table~\ref{tab:overall_performance}, \ourname outperforms all baseline methods in exception-handling code generation under both LLM configurations, achieving the highest CodeBLEU and Pass@1 scores.
Specifically, with GPT-4o, \ourname obtains a CodeBLEU score of 0.31 and a Pass@1 of 29\%, exceeding the strongest LLM-based baseline, RepoCoder (0.27 CodeBLEU, 25\% Pass@1), by 14.8\% and 4 percentage points, respectively.
A similar advantage is observed with DeepSeek-V3: \ourname attains a Pass@1 of 26\%, outperforming RepoCoder (23\%) and Direct-Prompting (10\%) by 3 and 16 percentage points, demonstrating that its design generalizes effectively beyond proprietary models.
In intent prediction (IntentAcc), \ourname also leads under GPT-4o (60.1\%) and DeepSeek-V3 (53.9\%), with substantial margins over all other methods.

The baselines exhibit clear limitations. ExAssist relies on static rule-based API–exception mappings and ignores repository context, resulting in low accuracy (9\% Pass@1). KPC merely re-throws caught exceptions and lacks strategic diversity. NexGen uses local context but misses repository-level information, reaching only 14\% Pass@1. Direct-Prompting produces generic code without grounding in exception-specific knowledge, while Seeker's multi-agent collaboration yields only marginal gains (15\%). RepoCoder achieves the strongest baseline performance by leveraging repository-wide context, yet its lack of explicit exception-aware reasoning limits further improvement. RepoCoder-RG1, which employs a simpler retrieval strategy, performs worse, underscoring the importance of precise context retrieval. 

To further examine performance differences, Figure~\ref{fig:rq1_venn} visualizes the overlap of correctly handled instances among representative methods under GPT-4o.
While all methods share a non-trivial common subset of solvable cases, \ourname covers the largest portion of unique correct instances, indicating that its improvements do not merely stem from solving easy or widely-addressed examples.
In particular, a substantial number of cases are correctly handled only by \ourname and are missed by other strong baselines such as RepoCoder and Seeker.
This suggests that \ourname can resolve exception-handling scenarios that require more than generic prompting or repository-level retrieval alone.
In contrast, baseline methods show considerable overlap with one another, reflecting limited diversity in their effective handling strategies.
%Overall, the Venn analysis complements the quantitative results in Table~\ref{tab:overall_performance}, confirming that \ourname not only raises average accuracy but also expands the coverage of solvable exception-handling cases through integrated knowledge.

These coverage-level differences provide concrete evidence that \ourname benefits from complementary knowledge sources rather than a single dominant factor. These consistent gains highlight the effectiveness of the three complementary knowledge channels in \ourname: (1) contextual call traces capturing repository-aware execution context, (2) exception-prone APIs narrowing the search space, and (3) handling patterns abstracted from real try-catch examples that guide generation. The ablation study (RQ3) further validates the contribution of each component.

%ExAssist exhibits relatively inferior performance due to its reliance on rule-based API-exception co-occurrence and its neglect of repository context, resulting in low coverage and accuracy. KPC assumes re-throwing as the only strategy, failing to reflect real-world diversity in handling. 
%Nexgen, a neural machine translation approach, performs slightly better than purely rule-based systems but lags behind retrieval-augmented and LLM-based methods. While it leverages both try-block and leading-code context, its reliance on shallow sequence modeling without repository-level or domain-specific knowledge constrains generalization.
%Direct-Prompting shows moderate performance, as LLMs can generalize generic handling logic but lack repository-level and exception-specific knowledge, limiting contextual accuracy.
%Seeker improves slightly via agent collaboration, but its lack of real-world handling pattern alignment affects reliability.
%RepoCoder performs best among baselines thanks to repository-level context, yet its omission of domain-specific knowledge (\emph{e.g.}, exception-prone APIs and handling patterns) limits further gains. Its variant, RepoCoder-RG1, with a simpler retrieval process, performs slightly worse, highlighting the value of precise and context-aware retrieval.

\smallskip\noindent\textbf{Answer to RQ1:} 
\ourname consistently outperforms all baselines in repository-aware exception handling, surpassing the strongest baseline, RepoCoder, by 14.8\%, 25.2\%, and 4\% (GPT-4o) / 3\% (DeepSeek-V3) in CodeBLEU, intent prediction accuracy, and Pass@1, respectively.

\subsection{Accuracy of Type Prediction (RQ2)}
\label{sec:RQ2}
\textbf{Setting.} 
We further evaluate exception type prediction accuracy of \ourname and other baselines on the \dataset benchmark. 
Considering Java's exception inheritance hierarchy, we measure prediction performance with three metrics at varying granularity levels:
(1) \textit{TypeAcc} evaluates exact matches between predicted and ground-truth exception types; (2) \textit{Parent TypeAcc} accepts predictions matching any superclass in the exception hierarchy; and (3) \textit{Child TypeAcc} considers matches with more specialized descendant exceptions as valid. 
Based on its strong performance in RQ1, we employ GPT-4o as the foundational model for this evaluation.

\begin{table}[t]
\centering
%\begin{threeparttable}
\centering
\caption{Accuracy of Exception Type Prediction with GPT-4o Model}
\label{tab:typeprediction}
\resizebox{\columnwidth}{!}{
\begin{tabular}{lccc}
\toprule
\textbf{Approach} & \textbf{TypeAcc} & \textbf{Parent TypeAcc} & \textbf{Child TypeAcc}  \\
\midrule
Direct-Prompting & 39.2\% & 40.4\% & 55.1\% \\
RepoCoder RG1 & 29.0\% & 29.5\% & 37.7\% \\
RepoCoder & 40.2\% & 40.3\% & 49.7\% \\
ExAssist & 27.2\% & 31.2\% & 35.2\% \\
Nexgen & 27.8\% & 32.0\% & 30.0 \% \\
KPC & 41.1\% & 44.3\% & 59.7\% \\
Seeker & 34.8\% & 36.0\% & 54.8\% \\
\ourname (ours) & \textbf{53.6\%} & \textbf{54.5\%} & \textbf{67.6\%} \\
\bottomrule
\end{tabular}
}
%    \begin{tablenotes}
%        \item * RepoCoder-RG1: single-round variant of RepoCoder. The best scores are in bold. 
%    \end{tablenotes}
%\end{threeparttable}  
\end{table}

\smallskip\noindent\textbf{Result.}
Table~\ref{tab:typeprediction} presents a performance comparison between our method and all baselines for exception type prediction. Overall, \ourname consistently achieves the best results across all metrics.
Specifically, \ourname achieves a TypeAcc of 53.6\%, exceeding the top baseline KPC (41.1\%) by 12.5\%, along with gains of 10.2\% and 7.9\% on Parent and Child TypeAcc, respectively. 

Existing methods commonly suffer from two key limitations: insufficient knowledge of exception types and a lack of repository-level context awareness.
KPC, for instance, restricts its attention to APIs within the local code block and relies heavily on the base model's built-in API knowledge, resulting in limited coverage.
Seeker incorporates certain prior knowledge from documentation and base models, yet it lacks exposure to real-world exception distributions and fails to account for broader call contexts.
Direct-Prompting depends entirely on the base model's generic prior knowledge without integrating any repository-specific signals.
Notably, Nexgen and ExAssist are outperformed by \ourname, with absolute TypeAcc improvements of 25.8 and 26.4 percentage points, respectively.
The performance gap arises because Nexgen learns exception knowledge from real-world repositories but overlooks long-range contextual dependencies, while ExAssist encodes exception knowledge through handcrafted rules derived from real projects yet remains context-agnostic.

%\ourname outperforms Direct-Prompting and ExAssist with absolute TypeAcc gains of 14.4 and 26.4 percentage points, respectively. The performance gap arises because Direct-Prompting operates solely on isolated code fragments, while ExAssist relies on predetermined templates that lack adaptive comprehension of repository structure.

In contrast, \ourname leverages a large-scale API-Exception mapping to model statistical associations, which is especially useful for APIs with sparse representation in individual repositories. Furthermore, by incorporating contextual call traces, our method integrates repository-wide context, which is essential for accurate exception type prediction since many exceptions originate from deep API invocations across files. This repository-aware reasoning enhances overall prediction accuracy while also explaining the consistently high Child TypeAcc performance. Even when the exact exception type cannot be determined, \ourname frequently identifies semantically relevant descendant types, demonstrating its advanced type inference capabilities.

\smallskip\noindent\textbf{Answer to RQ2:} 
\ourname significantly improves exception type prediction accuracy, achieving 53.6\% TypeAcc with an absolute improvement of 12 percentage points over the best-performing baseline.

\subsection{Contribution of Knowledge Components (RQ3)}
\label{sec:RQ3}
\textbf{Setting.} 
We conduct an ablation study to investigate the individual contributions of three core knowledge components, including contextual call traces, API-Exception mapping, and handling patterns.
In this study, we remove individual components one at a time while keeping others intact, then measure the performance impact on the resulting ablated variants. Throughout these experiments, we continue using GPT-4o as the base model and \dataset as the evaluation benchmark to maintain consistency with our main results, following the same metrics as in RQ1.

\begin{table*}[t]
\centering
\label{tab:ablation}
\begin{threeparttable}
\caption{Results of Ablation Study with GPT-4o Model}
%\resizebox{\columnwidth}{!}{
\begin{tabular}{l@{\hspace{6mm}}l@{\hspace{6mm}}l@{\hspace{6mm}}l}
\toprule
\textbf{Variants} & \textbf{Pass@1} & \textbf{CodeBLEU} & \textbf{IntentAcc} \\
\midrule
\ourname (Ours) & 29\% & 0.31 & 60.1\% \\
\quad w/o Contextual Call Traces & 23\% \scriptsize{($\downarrow$ 20.7\%)} & 0.29 \scriptsize{($\downarrow$ \: 6.5\%)} & 57.1\% \scriptsize{($\downarrow$ \: 5.0\%)} \\
\quad w/o API-Exception Mapping & \: 9\% \scriptsize{($\downarrow$ 67.0\%)} & 0.25  \scriptsize{($\downarrow$ 19.4\%)} & 48.3\% \scriptsize{($\downarrow$ 19.6\%)} \\
\quad w/o Exception Handling patterns  & 21\% \scriptsize{($\downarrow$ 27.6\%)} & 0.28 \scriptsize{($\downarrow$ \: 9.7\%)} & 48.8\% \scriptsize{($\downarrow$ 18.8\%)} \\
\bottomrule
\end{tabular}
%}
\end{threeparttable}
\end{table*}

\smallskip\noindent\textbf{Result.} 
The results are shown in Table~\ref{tab:ablation}. All three knowledge components contribute positively to the final performance. Removing any of them results in a significant degradation across all metrics, while their combination yields the most robust and context-aware exception handling generation. In particular, the overall Pass@1 score drops from 29\% to 23\%, 21\%, and 9\% when removing contextual call traces, handling patterns, and API-Exception mapping respectively, suggesting that each channel contributes to executable correctness, but the API-Exception mapping is indispensable for ensuring functional validity.

The extent of the performance loss varies per component, highlighting their different functionalities:
\begin{itemize}
    \item \textit{API-Exception Mapping}. This component proves the most critical. Its removal leads to the largest performance drop, with CodeBLEU decreasing by 19.4\%, IntentAcc by 19.6\%, and Pass@1 plummeting to only 9\%. Without knowledge of exception-prone APIs, the model fails to identify relevant exception types, which in turn impairs sample retrieval and pattern abstraction, causing broad performance decline.

    \item \textit{Contextual Call Traces}. Removing contextual call traces results in a noticeable performance decrease in CodeBLEU by 6.5\%, IntentACC by 5.0\%, and Pass@1 by 6 points (from 29\% to 23\%). This indicates that the contextual call trace and corresponding function signatures provide valuable semantic information essential for cross-procedural exception reasoning.

    \item \textit{Handling Patterns}. Replacing structured handling patterns with few-shot examples causes significant performance degradations, reducing CodeBLEU by 9.7\%, IntentACC by 18.8\%, and Pass@1 by 8 points (from 29\% to 21\%). This suggests that abstracted patterns provide reusable knowledge essential for generating generalizable code, beyond simple example imitation.
\end{itemize}

\smallskip\noindent\textbf{Answer to RQ3:} 
All three knowledge components are essential to \ourname. 
The API–Exception mapping is the most critical: removing it drops Pass@1 from 29\% to 9\%, showing it serves as the core mechanism for linking repository semantics to executable correctness. 
Contextual traces and handling patterns further enhance structural quality and semantic alignment.

\subsection{Impact of Retrieved Sample Number (RQ4)}
\label{sec:RQ4}

\textbf{Setting.} 
The exception handling patterns, a core knowledge component of \ourname, are abstracted from the top-$k$ most similar samples retrieved from the historical code corpus. 
Since pattern quality is highly sensitive to $k$, we evaluate its impact by varying $k$ from 0 to 30 to identify the optimal number of retrieved samples.
All other settings and evaluation metrics remain consistent with RQ4.

%\begin{wrapfigure}{r}{0.5\linewidth}
\begin{figure}[t]
\centerline{\includegraphics[width=0.9\columnwidth, trim=0 0 0 0, clip]{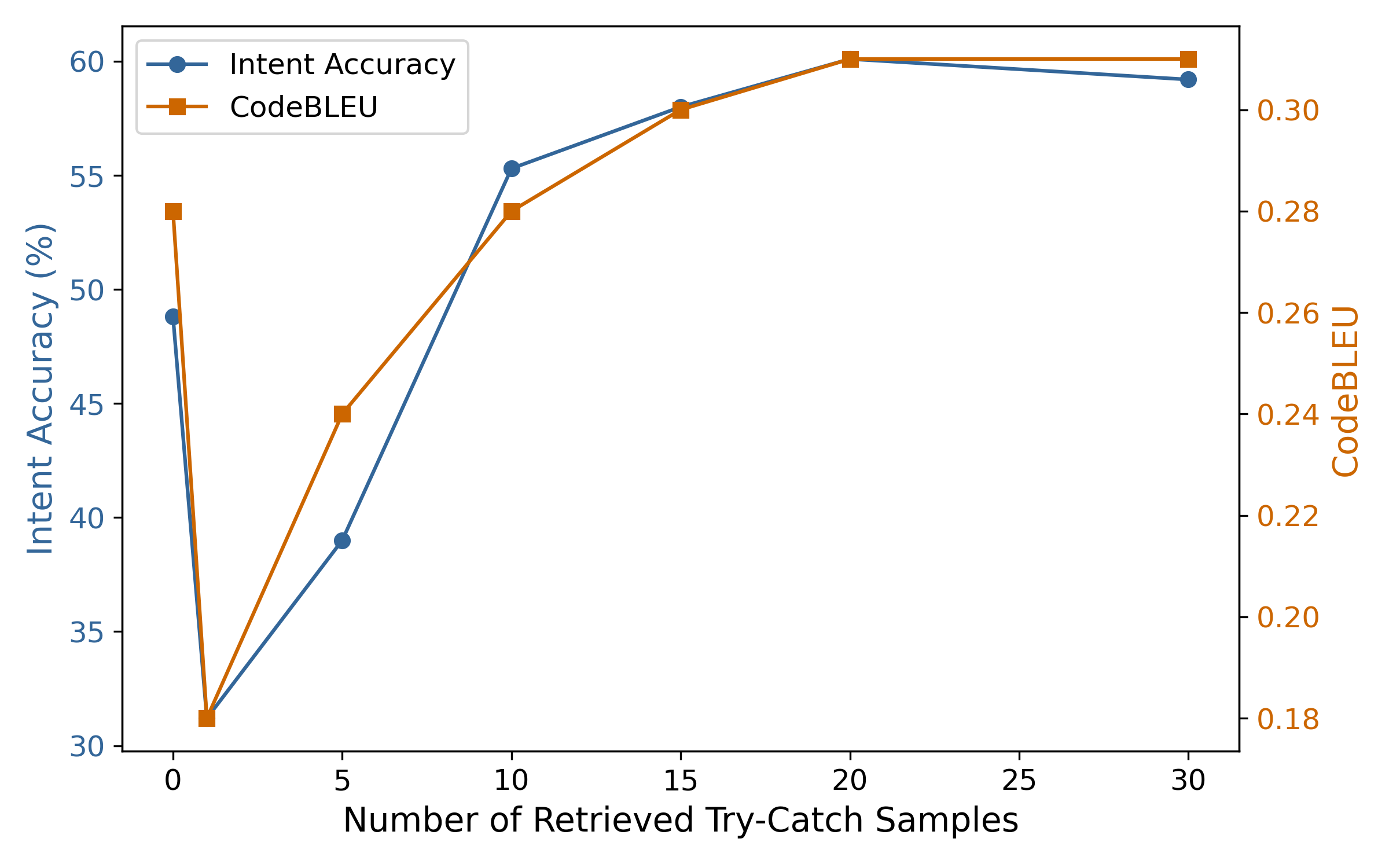}}
    \caption{Performance of \ourname under Different Numbers of Retrieved Try-Catch Samples.} 
    \label{fig:count}
\end{figure}    
%\end{wrapfigure}

\smallskip\noindent\textbf{Result.} 
As shown in Figure~\ref{fig:count}, using only 1 or 5 samples results in a noticeable performance drop compared to the zero-sample baseline. This performance reduction implies that small sample sizes may yield noisy or biased patterns that mislead the subsequent code generation. 

% Performance improves substantially with 10 or more samples, indicating that greater diversity supports more robust pattern abstraction. Optimal performance occurs around 20 samples. Beyond this point, intent accuracy slightly declines at $k = 30$, likely due to the inclusion of irrelevant or conflicting patterns. 

% In addition, the trend of \textit{Pass@1} shows a steady improvement as $k$ increases. When $k$ is small (1–5), the retrieved samples fail to provide sufficient contextual grounding, leading to handlers that rarely pass functional validation. As $k$ increases, the improvement in \textit{Pass@1} indicates that richer contextual diversity enhances the model's ability to generate semantically correct and executable handling logic. The metric peaks around $k=30$, suggesting that even when textual similarity and intent alignment have stabilized, functional correctness continues to benefit from the expanded sample set.

Performance improves as $k$ increases, with substantial gains once at least 10 samples are retrieved, indicating that greater diversity enables more robust pattern abstraction. Performance peaks around $k=20$, after which intent accuracy slightly declines (e.g., at $k=30$), likely due to noise from irrelevant or conflicting samples. 
Meanwhile, \textit{Pass@1} steadily rises with larger $k$, showing that richer contextual diversity helps produce more executable and semantically correct handlers; however, returns diminish as similarity saturation leads to marginal or even negative effects beyond the optimal range.

These results reveal a key trade-off between abstraction reliability and sample size, where both insufficient and excessive retrieved examples can compromise generation quality through underfitting and overfitting respectively.

\smallskip\noindent\textbf{Answer to RQ4:} 
\ourname achieves optimal performance with around 20 retrieved samples. 
Fewer than 10 samples lack sufficient diversity, while larger sets yield diminishing returns—though \textit{Pass@1} continues to improve slightly up to 30, reflecting gains in functional robustness.

\subsection{Case Study}
\label{sec:CaseStudy}

\begin{figure*}[t]
\centerline{\includegraphics[width=1.7\columnwidth, trim=0 0 0 0, clip]{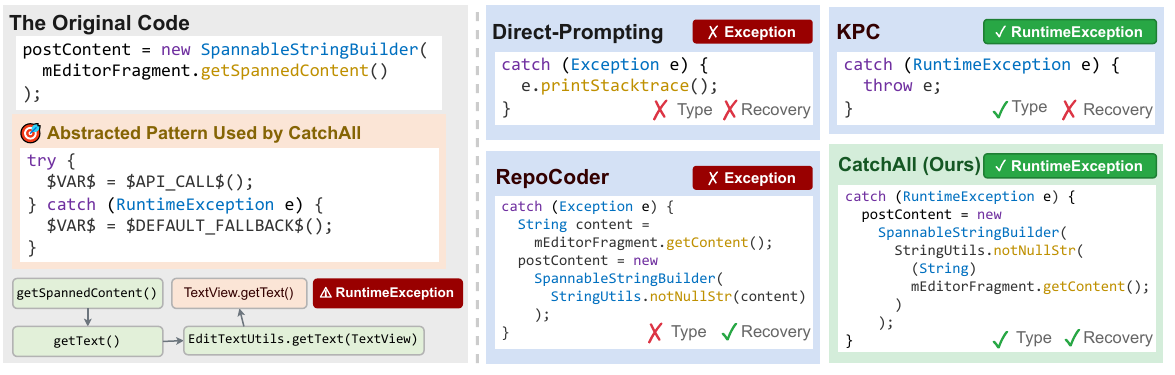}}
\caption{Examples of Catch Blocks Generated by Different Approaches for a WordPress Android Code Snippet.}
\label{fig:case-study}
\end{figure*}

To provide an intuitive demonstration of \ourname's effectiveness, we present a representative case from the WordPress Android project, as shown in Figure~\ref{fig:case-study}. The code instantiates a \verb|SpannableStringBuilder| using content from the post editor. Tracing \verb|getSpannedContent()| reveals a call chain across three files, originating from the Android SDK's \verb|textView.getText()|. The complete trace includes four methods: \texttt{getSpannedContent() → getText() → EditTextUtils.getText(TextView) → TextView.getText()}. This chain risks a \verb|RuntimeException| if \verb|textView| is dereferenced while null, a challenging scenario that necessitates both cross-file analysis and exception knowledge to handle correctly.

Existing approaches exhibit significant limitations in addressing such cases. Direct-Prompting fails to predict the specific exception type, defaulting to a generic \verb|Exception| catch block that merely prints the stack trace. Lacking repository-aware context and exception-specific knowledge, the language model cannot accurately infer the error's root cause or produce meaningful recovery behavior. 
RepoCoder performs relatively better in generating meaningful recovery logic and method calls, accurately reconstructing fallback behavior. However, it still fails to predict the correct exception type and instead falls back to the generic \verb|Exception|, reflecting its limited grasp of API-specific exception semantics. Moreover, its generation process is time-consuming, taking over 40 seconds compared to under 5 seconds for other methods. 
Meanwhile, KPC consistently re-throws exceptions due to its throw-centric rewriting strategy, which does not support alternative handling strategies such as recovery or fallback mechanisms. This limitation considerably narrows its applicability in real-world scenarios where graceful error recovery is essential.

In contrast, \ourname accurately predicts the exception type and generates appropriate recovery logic. This is enabled by the synergy of three knowledge channels: (1) \textit{Contextual call traces} locate the exception risk at \verb|TextView.| \verb|getText()| via cross-file analysis; (2) \textit{API-Exception mapping} links this risk to \verb|RuntimeException| using learned API behavior; (3) \textit{Exception handling patterns} supply a reusable template from prior \verb|try-catch| blocks in the repository. The abstracted pattern specifies that when a method throws \verb|RuntimeException|, the variable should be assigned a safe fallback value via an alternative method. Accordingly, \ourname replaces the risky \verb|getSpannedContent()| call with \verb|StringUtils.| \verb|notNullStr((String) mEditorFragment.getContent())|, a robust, repository-aligned fallback. This case confirms that integrating call traces, API-exception mappings, and handling patterns allows precise exception prediction and context-aware code generation.

\section{Discussion}

\subsection{Why Focus on \texttt{Try-Catch}?}

\ourname focuses on synthesizing repository-aware \texttt{try-catch} blocks as its primary exception-handling mechanism. This design choice aligns with established practices in exception-handling research~\cite{fetzer2004automatic, zhong2022exception}, where the construction of \texttt{try-catch} blocks is consistently treated as a fundamental task. Several factors motivate this focus:

\begin{itemize}
\item \textbf{Prevalence in practice.} The \texttt{try-catch} construct is the most common explicit exception-handling paradigm in Java and Android ecosystems. Our analysis of 3,200 Android repositories confirms its dominance, revealing that over 85\% of explicit exception-handling code utilizes \texttt{try-catch} blocks.

\item \textbf{Timeliness of exception handling.} Catching exceptions where they occur with \texttt{try-catch} ensures that the program can gracefully handle the error and safely continue or terminate, while also making the code more self-contained and easier to reason about by encapsulating error-handling logic near its source. Therefore, we advocate for intercepting and handling the exception flow as early as appropriate, proactively building robustness and improving maintainability in the process.

\item \textbf{Semantic clarity.} By designating specific exception types, handling strategies, and recovery logic, \texttt{Try-catch} blocks establish clear semantic boundaries between the normal execution flow and error-handling logic. This explicit separation enhances structural clarity, which in turn facilitates both static program analysis and LLM-assisted tasks such as exception type prediction and exception-handling code generation.

\end{itemize}

We acknowledge that practical error management encompasses strategies beyond \texttt{try-catch}. Our current approach does not synthesize complementary techniques such as preventive handling~\cite{10.1145/2522920.2522923} (e.g., using preconditions to avoid exceptions), logging and observability~\cite{10.1145/2591062.2591175} (for post-mortem analysis), exception propagation and transformation~\cite{10.1145/2652483}, or structured resource management (e.g., \texttt{try-}\texttt{with-} \texttt{resources}). Extending \ourname to incorporate a broader spectrum of error-handling patterns represents a valuable direction for future work, which would enhance the tool's practical utility and coverage of real-world practices.

\subsection{What Is the Computational Efficiency of \ourname?}

We analyze the computational efficiency and practical scalability of \ourname. Its cost structure comprises a one-time, upfront investment and a minimal marginal cost per synthesis query:

\begin{itemize}
\item \textbf{Offline knowledge construction.} Extracting the API-Exception mapping from our corpus of 3,233 repositories requires approximately 97 hours of computation. This is a one-time cost incurred to build the foundational knowledge base and is amortized over all subsequent uses.

\item \textbf{Per-query inference cost.} For each query during inference, the call-trace analysis averages 4.6 seconds. The subsequent LLM inference typically consumes about 1,800 input tokens and generates approximately 50 output tokens.
\end{itemize}

This cost profile ensures scalability. The upfront investment enables practical deployment in environments like IDE plugins, while the low per-query overhead supports real-time, interactive developer assistance.

%\subsection{Why is \ourname Effective?}
%The effectiveness of \ourname stems from its ability to enrich LLMs with deep knowledge of repository context. It harnesses the powerful representation and reasoning capabilities of LLMs in code understanding and generation, and enhances them by integrating three key sources of knowledge—execution context, exception-prone APIs, and handling patterns—into a unified framework.

%Unlike rule-based methods such as ExAssist, \ourname adopts a data-driven approach that automatically learns exception handling patterns from large-scale repositories. This eliminates the need for manual rule crafting and enhances the model's adaptability and generalization across diverse projects.

%Compared to knowledge-driven approaches like KPC and Seeker, \ourname incorporates contextual call traces to capture rich cross-file execution context, which is often crucial for accurate exception handling.

%Furthermore, while general-purpose code generation models like RepoCoder focus broadly on code synthesis, \ourname specifically targets exception-related knowledge by mining real-world repositories for common handling patterns and identifying APIs that are prone to exceptions.

%By integrating these knowledge sources, \ourname significantly improves the accuracy of exception handling while narrowing the solution space.

\subsection{Threats to Validity}

We identify the following limitations and potential threats to the validity of our study:

\textit{Data leakage:}
Our method employs LLMs to generate exception-handling code. As these models scale, their training data coverage expands, increasing the risk of exposure to samples from our test dataset. To mitigate this issue, we apply time-based filtering techniques to minimize training-test data overlap.
We further conduct an exact-match verification against CodeSearchNet, a representative corpus commonly included in LLM pretraining, finding only 4.4\% overlap (143 out of 3,233 repositories).
However, even with these safeguards, a residual risk of data leakage persists.

\textit{Evaluation of generated code:}
We primarily evaluate \ourname on our large-scale \dataset benchmark using CodeBLEU and intent prediction accuracy. To assess functional correctness, we additionally construct an executable benchmark (\datasetexec) and evaluate it using GPT-4o. However, due to constraints in time and computational resources, the current version of \datasetexec is limited to four repositories. Future work will expand this benchmark to include more repositories and perform more comprehensive, test-based evaluations across a wider range of models.

\section{Conclusion}
\label{sec:conclusion}
This paper introduces \ourname, a novel LLM-based approach for repository-aware exception handling. 
\ourname implements a knowledge augmentation framework to extract and integrate contextual call traces, exception-prone APIs, and handling patterns, enhancing LLMs to synthesize accurate \texttt{try-catch} blocks within repository context. 
The results demonstrate that \ourname outperforms state-of-the-art baselines by a significant margin, achieving superior performance in both exception type prediction and handling code generation.
In the future, we will develop a multilingual, repository-aware benchmarking framework with test validation capabilities to support multi-faceted evaluations and advance the state of research in repository-aware exception handling.
%We will also explore integrating dynamic runtime information and exception stack traces to further enhance prediction accuracy and practical applicability.

Source code and dataset to reproduce our work are available at
\url{https://github.com/q4x3/CatchAll}.

%\section*{Acknowledgments}

\appendix
%\section{My Appendix}
%Appendix sections are coded under \verb+\appendix+.

%\verb+\printcredits+ command is used after appendix sections to list author credit taxonomy contribution roles tagged using \verb+\credit+ in frontmatter.

\printcredits

%% Loading bibliography style file
%\bibliographystyle{model1-num-names}
%\bibliographystyle{cas-model2-names}
\balance
\bibliographystyle{elsarticle-num}

% Loading bibliography database
\bibliography{main.bib}

@inproceedings{ZhangWZSPL20,
  author       = {Jian Zhang and
                  Xu Wang and
                  Hongyu Zhang and
                  Hailong Sun and
                  Yanjun Pu and
                  Xudong Liu},
  title        = {Learning to Handle Exceptions},
  booktitle    = {35th {IEEE/ACM} International Conference on Automated Software Engineering
                  ({ASE}), Melbourne, Australia, September 21-25, 2020},
  pages        = {29--41},
  publisher    = {{IEEE}},
  year         = {2020}
}

@inproceedings{cabral2007exception,
  title={Exception handling: A field study in {Java} and. net},
  author={Cabral, Bruno and Marques, Paulo},
  booktitle={Proc. ECOOP},
  pages={151--175},
  year={2007}
}

@inproceedings{sena2016understanding,
  title={Understanding the exception handling strategies of {J}ava libraries: An empirical study},
  author={Sena, Dem{\'o}stenes and Coelho, Roberta and Kulesza, Uir{\'a} and Bonif{\'a}cio, Rodrigo},
  booktitle={Proc. MSR},
  pages={212--222},
  year={2016}
}

@inproceedings{jana2016automatically,
  title={Automatically detecting error handling bugs using error specifications},
  author={Jana, Suman and Kang, Yuan Jochen and Roth, Samuel and Ray, Baishakhi},
  booktitle={Proc. USENIX Security },
  pages={345--362},
  year={2016}
}

@inproceedings{liang2025can,
  title={Can Language Models Replace Programmers for Coding? REPOCOD Says ‘Not Yet’},
  author={Liang, Shanchao and Jiang, Nan and Hu, Yiran and Tan, Lin},
  booktitle={Proc. ACL},
  pages={24698--24717},
  year={2025}
}

@inproceedings{nguyen2019recommending,
  title={Recommending exception handling code},
  author={Nguyen, Tam and Vu, Phong and Nguyen, Tung},
  booktitle={Proc. ICSME},
  pages={390--393},
  year={2019}
}

@inproceedings{tam2020,
  title={Code recommendation for exception handling},
  author={Tam Nguyen and Phong Vu and Tung Nguyen},
  booktitle={Proc. ESEC/FSE},
  pages={1027--1038},
  year={2020}
}

@article{jo2004uncaught,
  title={An uncaught exception analysis for {J}ava},
  author={Jo, Jang-Wu and Chang, Byeong-Mo and Yi, Kwangkeun and Choe, Kwang-Moo},
  journal={Journal of systems and software},
  volume={72},
  number={1},
  pages={59--69},
  year={2004}
}

@inproceedings{rahman2014use,
  title={On the use of context in recommending exception handling code examples},
  author={Rahman, Mohammad Masudur and Roy, Chanchal K},
  booktitle={Proc. SCAM},
  pages={285--294},
  year={2014}
}

@inproceedings{zhong2022exception,
  title={Which Exception Shall We Throw?},
  author={Zhong, Hao},
  booktitle={Proc. ASE},
  pages={1--12},
  year={2022}
}

@inproceedings{xu2021exception,
Author={Lin Xu and Hao Zhong},
Booktitle={Proc. ICPC},
title={Detecting inconsistent thrown exceptions},
year={2021},
pages={391--395},
}

@inproceedings{jia2021handle,
  title={Where to handle an exception? Recommending exception handling locations from a global perspective},
  author={Jia, Xiangyang and Chen, Songqiang and Zhou, Xingqi and Li, Xintong and Yu, Run and Chen, Xu and Xuan, Jifeng},
  booktitle={Proc. ICPC},
  pages={369--380},
  year={2021},
}

@article{fetzer2004automatic,
  title={Automatic detection and masking of nonatomic exception handling},
  author={Fetzer, Christof and Felber, Pascal and Hogstedt, Karin},
  journal={IEEE Transactions on Software Engineering},
  volume={30},
  number={8},
  pages={547--560},
  year={2004}
}

@inproceedings{acharya09:mining,
	Author = {Mithun Acharya and Tao Xie},
	Booktitle = {Proc. FASE},
	Pages = {370--384},
	Title = {Mining {API} Error-Handling Specifications from Source Code},
	Year = {2009}}

@inproceedings{cacho2014does,
  title={How does exception handling behavior evolve? an exploratory study in {J}ava and {C}\# applications},
  author={Cacho, N{\'e}lio and Barbosa, Eiji Adachi and Araujo, Juliana and Pranto, Frederico and Garcia, Alessandro and Cesar, Thiago and Soares, Eliezio and Cassio, Arthur and Filipe, Thomas and Garcia, Israel},
  booktitle={Proc. ICSME},
  pages={31--40},
  year={2014}
}

@inproceedings{chen2019understanding,
  title={Understanding Exception-Related Bugs in Large-Scale Cloud Systems},
  author={Chen, Haicheng and Dou, Wensheng and Jiang, Yanyan and Qin, Feng},
  booktitle={Proc. ASE},
  pages={339--351},
  year={2019},
}

@inproceedings{fan2018large,
  title={Large-scale analysis of framework-specific exceptions in {Android} {A}pps},
  author={Fan, Lingling and Su, Ting and Chen, Sen and Meng, Guozhu and Liu, Yang and Xu, Lihua and Pu, Geguang and Su, Zhendong},
  booktitle={Proc. ICSE},
  pages={408--419},
  year={2018}
}

@article{shah2010understanding,
  title={Understanding exception handling: Viewpoints of novices and experts},
  author={Shah, Hina and Gorg, Carsten and Harrold, Mary Jean},
  journal={IEEE Transactions on Software Engineering},
  volume={36},
  number={2},
  pages={150--161},
  year={2010},
}

@article{koopman2000exception,
  title={The exception handling effectiveness of POSIX operating systems},
  author={Koopman, Phil and DeVale, John},
  journal={IEEE Transactions on Software Engineering},
  volume={26},
  number={9},
  pages={837--848},
  year={2000},
}

@inproceedings{bruntink2006discovering,
  title={Discovering faults in idiom-based exception handling},
  author={Bruntink, Magiel and Van Deursen, Arie and Tourw{\'e}, Tom},
  booktitle={Proc. ICSE},
  pages={242--251},
  year={2006}
}

@inproceedings{asaduzzaman2016developers,
  title={How developers use exception handling in {Java}?},
  author={Asaduzzaman, Muhammad and Ahasanuzzaman, Muhammad and Roy, Chanchal K and Schneider, Kevin A},
  booktitle={Proc. of MSR},
  pages={516--519},
  year={2016}
}

@article{coelho2017exception,
  title={Exception handling bug hazards in {Android}},
  author={Coelho, Roberta and Almeida, Lucas and Gousios, Georgios and Van Deursen, Arie and Treude, Christoph},
  journal={Empirical Software Engineering},
  volume={22},
  number={3},
  pages={1264--1304},
  year={2017}
}

@inproceedings{Zhong22a,
  author       = {Hao Zhong},
  title        = {Which Exception Shall We Throw?},
  booktitle    = {37th {IEEE/ACM} International Conference on Automated Software Engineering,
                  {ASE} 2022, Rochester, MI, USA, October 10-14, 2022},
  pages        = {116:1--116:12},
  publisher    = {{ACM}},
  year         = {2022},
}

@article{seeker,
  author       = {Xuanming Zhang and
                  Yuxuan Chen and
                  Yuan Yuan and
                  Minlie Huang},
  title        = {Seeker: Enhancing Exception Handling in Code with LLM-based Multi-Agent
                  Approach},
  journal      = {CoRR},
  volume       = {abs/2410.06949},
  year         = {2024},
}

@inproceedings{ExAssist,
  author       = {Tam Nguyen and
                  Phong Vu and
                  Tung Nguyen},
  booktitle    = {{ESEC/FSE} '20: 28th {ACM} Joint European Software Engineering Conference
                  and Symposium on the Foundations of Software Engineering, Virtual
                  Event, USA, November 8-13, 2020},
  pages        = {1027--1038},
  publisher    = {{ACM}},
  year         = {2020},
}

@inproceedings{KPC,
  author       = {Xiaoxue Ren and
                  Xinyuan Ye and
                  Dehai Zhao and
                  Zhenchang Xing and
                  Xiaohu Yang},
  title        = {From Misuse to Mastery: Enhancing Code Generation with Knowledge-Driven {AI} Chaining},
  booktitle    = {38th {IEEE/ACM} International Conference on Automated Software Engineering,
                  {ASE} 2023, Luxembourg, September 11-15, 2023},
  pages        = {976--987},
  publisher    = {{IEEE}},
  year         = {2023},

}

@inproceedings{RepoCoder,
  author       = {Fengji Zhang and
                  Bei Chen and
                  Yue Zhang and
                  Jacky Keung and
                  Jin Liu and
                  Daoguang Zan and
                  Yi Mao and
                  Jian{-}Guang Lou and
                  Weizhu Chen},
  title        = {RepoCoder: Repository-Level Code Completion Through Iterative Retrieval
                  and Generation},
  booktitle    = {Proceedings of the 2023 Conference on Empirical Methods in Natural
                  Language Processing, {EMNLP} 2023, Singapore, December 6-10, 2023},
  pages        = {2471--2484},
  year         = {2023},

}

@article{codebleu2020ren,
  author       = {Shuo Ren and
                  Daya Guo and
                  Shuai Lu and
                  Long Zhou and
                  Shujie Liu and
                  Duyu Tang and
                  Neel Sundaresan and
                  Ming Zhou and
                  Ambrosio Blanco and
                  Shuai Ma},
  title        = {CodeBLEU: a Method for Automatic Evaluation of Code Synthesis},
  journal      = {CoRR},
  volume       = {abs/2009.10297},
  year         = {2020}
}

@inproceedings{IntentClassification,
  author       = {Soham Parikh and
                  Mitul Tiwari and
                  Prashil Tumbade and
                  Quaizar Vohra},
  title        = {Exploring Zero and Few-shot Techniques for Intent Classification},
  booktitle    = {{ACL} 2023, Toronto, Canada,
                  July 9-14, 2023},
  pages        = {744--751},

  year         = {2023},

}

@article{melo2019unveiling,
  title={Unveiling Exception Handling Guidelines Adopted by {J}ava Developers},
  author={Melo, Hugo and Coelho, Roberta and Treude, Christoph},
  journal={arXiv preprint arXiv:1901.08718},
  year={2019},
}

@article{padua2017revisiting,
  title={Revisiting Exception Handling Practices with Exception Flow Analysis},
  author={de Pádua, Guilherme B. and Shang, Weiyi},
  journal={arXiv preprint arXiv:1708.00817},
  year={2017},
}

@article{SilvaVMGR24,
  author       = {Ant{\^{o}}nio da Silva and
                  Renan Gomes Vieira and
                  Diego P. P. Mesquita and
                  Jo{\~{a}}o Paulo Pordeus Gomes and
                  Lincoln S. Rocha},
  title        = {Towards automatic labeling of exception handling bugs: {A} case study
                  of 10 years bug-fixing in Apache Hadoop},
  journal      = {Empir. Softw. Eng.},
  volume       = {29},
  number       = {4},
  pages        = {85},
  year         = {2024},

}

@article{CodeHunter, 
title={{BERT}-Based Code Learning for Exception Localization and Type Prediction}, 
volume={39}, 
number={1}, 
journal={Proceedings of the AAAI Conference on Artificial Intelligence}, 
author={Zhang, Chongyu and Tao, Qiping and Chen, Liangyu and Zhang, Min}, 
year={2025}, month={Apr.}, pages={1040-1047} }

@inproceedings{eha,
author = {Zhang, Hao and Luo, Ji and Hu, Mengze and Yan, Jun and Zhang, Jian and Qiu, Zongyan},
title = {Detecting Exception Handling Bugs in C++ Programs},
year = {2023},
isbn = {9781665457019},
publisher = {IEEE Press},
booktitle = {Proceedings of the 45th International Conference on Software Engineering},
pages = {1084–1096},
numpages = {13},
location = {Melbourne, Victoria, Australia},
series = {ICSE '23}
}

@inproceedings{ExceRef,
author = {Zhang, Yang and Xue, Yuan},
title = {ExceRef: Automatically Refactoring for Exception Handling},
year = {2024},
booktitle = {Proceedings of the 15th Asia-Pacific Symposium on Internetware},
pages = {239–248},
numpages = {10},
location = {Macau, China},
series = {Internetware '24}
}

@INPROCEEDINGS{Neurex,
  author={Cai, Yuchen and Yadavally, Aashish and Mishra, Abhishek and Montejo, Genesis and Nguyen, Tien N.},
  booktitle={2024 IEEE/ACM 46th International Conference on Software Engineering (ICSE)}, 
  title={Programming Assistant for Exception Handling with {CodeBERT}}, 
  year={2024},
  pages={1146-1158},
}

@inproceedings{Kery2016,
author = {Kery, Mary Beth and Le Goues, Claire and Myers, Brad A.},
title = {Examining programmer practices for locally handling exceptions},
year = {2016},
booktitle = {Proceedings of the 13th International Conference on Mining Software Repositories},
pages = {484–487},
numpages = {4},

location = {Austin, Texas},
series = {MSR '16}
}

@INPROCEEDINGS{7961532,
  author={De Pádua, Guilherme Bicalho and Shang, Weiyi},
  booktitle={2017 IEEE/ACM 25th International Conference on Program Comprehension (ICPC)}, 
  title={Studying the Prevalence of Exception Handling Anti-Patterns}, 
  year={2017},

  pages={328-331},
}

@INPROCEEDINGS{10174045,
  author={Oliveira, Anderson and Correia, João and Sousa, Leonardo and Assunção, Wesley K. G. and Coutinho, Daniel and Garcia, Alessandro and Oizumi, Willian and Barbosa, Caio and Uchôa, Anderson and Pereira, Juliana Alves},
  booktitle={2023 IEEE/ACM 20th International Conference on Mining Software Repositories (MSR)}, 
  title={Don’t Forget the Exception! : Considering Robustness Changes to Identify Design Problems}, 
  year={2023},
  pages={417-429},

}

@inproceedings {ExChain,
author = {Ao Li and Shan Lu and Suman Nath and Rohan Padhye and Vyas Sekar},
title = {{ExChain}: Exception Dependency Analysis for Root Cause Diagnosis},
booktitle = {21st USENIX Symposium on Networked Systems Design and Implementation (NSDI 24)},
year = {2024},
pages = {2047--2062},
}

@INPROCEEDINGS{9825827,
  author={Li, Rongfan and Chen, Bihuan and Zhang, Fengyi and Sun, Chao and Peng, Xin},
  booktitle={2022 IEEE International Conference on Software Analysis, Evolution and Reengineering (SANER)}, 
  title={Detecting Runtime Exceptions by Deep Code Representation Learning with Attention-Based Graph Neural Networks}, 
  year={2022},
  pages={373-384},
}

@INPROCEEDINGS{9462986,
  author={Jia, Xiangyang and Chen, Songqiang and Zhou, Xingqi and Li, Xintong and Yu, Run and Chen, Xu and Xuan, Jifeng},
  booktitle={2021 IEEE/ACM 29th International Conference on Program Comprehension (ICPC)}, 
  title={Where to Handle an Exception? Recommending Exception Handling Locations from a Global Perspective}, 
  year={2021},
  pages={369-380},
}

@INPROCEEDINGS{9978172,
  author={Marcilio, Diego and Furia, Carlo A.},
  booktitle={2022 IEEE International Conference on Software Maintenance and Evolution (ICSME)}, 
  title={What Is Thrown? Lightweight Precise Automatic Extraction of Exception Preconditions in Java Methods}, 
  year={2022},
  pages={340-351},
}

@inproceedings{3608132,
author = {Peng, Yun and Gao, Shuzheng and Gao, Cuiyun and Huo, Yintong and Lyu, Michael},
title = {Domain Knowledge Matters: Improving Prompts with Fix Templates for Repairing Python Type Errors},
year = {2024},
booktitle = {Proceedings of the IEEE/ACM 46th International Conference on Software Engineering},
articleno = {4},
numpages = {13},
location = {Lisbon, Portugal},
series = {ICSE '24}
}

@INPROCEEDINGS{5070548,
  author={Thummalapenta, Suresh and Xie, Tao},
  booktitle={2009 IEEE 31st International Conference on Software Engineering}, 
  title={Mining exception-handling rules as sequence association rules}, 
  year={2009},

  pages={496-506},

}

@article{10.1145/2522920.2522923,
author = {Chang, Herv\'{e} and Mariani, Leonardo and Pezz\`{e}, Mauro},
title = {Exception handlers for healing component-based systems},
year = {2013},
issue_date = {October 2013},
publisher = {Association for Computing Machinery},
address = {New York, NY, USA},
volume = {22},
number = {4},
issn = {1049-331X},
url = {https://doi.org/10.1145/2522920.2522923},
doi = {10.1145/2522920.2522923},
journal = {ACM Trans. Softw. Eng. Methodol.},
month = oct,
articleno = {30},
numpages = {40}
}

@inproceedings{10.1145/2591062.2591175,
author = {Fu, Qiang and Zhu, Jieming and Hu, Wenlu and Lou, Jian-Guang and Ding, Rui and Lin, Qingwei and Zhang, Dongmei and Xie, Tao},
title = {Where do developers log? an empirical study on logging practices in industry},
year = {2014},
isbn = {9781450327688},
publisher = {Association for Computing Machinery},
address = {New York, NY, USA},
url = {https://doi.org/10.1145/2591062.2591175},
doi = {10.1145/2591062.2591175},
booktitle = {Companion Proceedings of the 36th International Conference on Software Engineering},
pages = {24–33},
numpages = {10},
location = {Hyderabad, India},
series = {ICSE Companion 2014}
}

@article{10.1145/2652483,
author = {Zhang, Pingyu and Elbaum, Sebastian},
title = {Amplifying Tests to Validate Exception Handling Code: An Extended Study in the Mobile Application Domain},
year = {2014},
issue_date = {August 2014},
publisher = {Association for Computing Machinery},
address = {New York, NY, USA},
volume = {23},
number = {4},
issn = {1049-331X},
url = {https://doi.org/10.1145/2652483},
doi = {10.1145/2652483},
journal = {ACM Trans. Softw. Eng. Methodol.},
month = sep,
articleno = {32},
numpages = {28}
}

@misc{chen2021evaluatinglargelanguagemodels,
      title={Evaluating Large Language Models Trained on Code}, 
      author={Mark Chen and Jerry Tworek and Heewoo Jun and Qiming Yuan and Henrique Ponde de Oliveira Pinto and Jared Kaplan and Harri Edwards and Yuri Burda and Nicholas Joseph and Greg Brockman and Alex Ray and Raul Puri and Gretchen Krueger and Michael Petrov and Heidy Khlaaf and Girish Sastry and Pamela Mishkin and Brooke Chan and Scott Gray and Nick Ryder and Mikhail Pavlov and Alethea Power and Lukasz Kaiser and Mohammad Bavarian and Clemens Winter and Philippe Tillet and Felipe Petroski Such and Dave Cummings and Matthias Plappert and Fotios Chantzis and Elizabeth Barnes and Ariel Herbert-Voss and William Hebgen Guss and Alex Nichol and Alex Paino and Nikolas Tezak and Jie Tang and Igor Babuschkin and Suchir Balaji and Shantanu Jain and William Saunders and Christopher Hesse and Andrew N. Carr and Jan Leike and Josh Achiam and Vedant Misra and Evan Morikawa and Alec Radford and Matthew Knight and Miles Brundage and Mira Murati and Katie Mayer and Peter Welinder and Bob McGrew and Dario Amodei and Sam McCandlish and Ilya Sutskever and Wojciech Zaremba},
      year={2021},
      eprint={2107.03374},
      archivePrefix={arXiv},
      primaryClass={cs.LG},
      url={https://arxiv.org/abs/2107.03374}, 
}

@misc{wordpress_android_repo,
  author = {{wordpress-mobile}},
  title = {{WordPress-Android}},
  howpublished = {\url{https://github.com/wordpress-mobile/WordPress-Android}},
  year = {2025},
}

\end{document}